\documentclass[10pt, conference, letterpaper]{IEEEtran}
\IEEEoverridecommandlockouts
\usepackage{booktabs} 
\usepackage{amsmath}
\usepackage[english]{babel}
\usepackage[latin1]{inputenc}
\usepackage{amssymb}
\usepackage[ruled]{algorithm2e}
\usepackage{algpseudocode}
\usepackage{bm}
\usepackage{amsmath}
\usepackage{graphicx}
\usepackage{epsfig}
\usepackage{url}
\usepackage{psfrag}
\usepackage{balance}
\usepackage{pstricks}
\usepackage{times}
\usepackage{array}
\usepackage{subfigure}
\usepackage{float}
\usepackage{multirow}
\usepackage{mathtools}

\newtheorem{theorem}{\textbf{Theorem}}
\newenvironment{pf}{\noindent{\textbf{Proof.} }}{\hfill $\square$\medskip}
\algnewcommand\algorithmicforeach{\textbf{for each}}

\begin{document}
\title{Utilizing Dynamic Properties of Sharing Bits and Registers to Estimate User Cardinalities over Time
\thanks{\textsuperscript{*}Peng Jia, Jing Tao and Xiaohong Guan are corresponding authors.}
}
\author{
      \fontsize{11}{11}\selectfont
      Pinghui Wang$^{1,2}$, Peng Jia$^{1}$, Xiangliang Zhang$^{3}$, Jing Tao$^{1,2,4}$, Xiaohong Guan$^{2,1,5}$, Don Towsley$^{6}$\\
      $^{1}$MOE Key Laboratory for Intelligent Networks and Network Security, Xi'an Jiaotong University, China\\
      $^{2}$Shenzhen Research Institute of Xi'an Jiaotong University, Shenzhen, China\\
      $^{3}$King Abdullah University of Science and Technology, Thuwal, SA \\
      $^{4}$Zhejiang Research Institute of Xi'an Jiaotong University, Hangzhou, China\\
      $^{5}$Department of Automation and NLIST Lab, Tsinghua University, Beijing, China\\
      $^{6}$School of Computer Science, University of Massachusetts Amherst, MA, USA\\
      Email: \{phwang, jtao, xhguan, pengjia\}@sei.xjtu.edu.cn, xiangliang.zhang@kaust.edu.sa,\\
      towsley@cs.umass.edu\\
}

\maketitle

\begin{abstract}
Online monitoring user cardinalities (or degrees) in graph streams is fundamental for many applications.
For example in a bipartite graph representing user-website visiting activities, user cardinalities (the number of distinct visited websites) are monitored to report network anomalies.
These real-world graph streams may contain user-item duplicates and have a huge number of distinct user-item pairs,
therefore, it is infeasible to exactly compute user cardinalities when memory and computation resources are limited.
Existing methods are designed to approximately estimate user cardinalities, whose accuracy highly depends on parameters that are not easy to set.
Moreover, these methods cannot provide anytime-available estimation, as the  user cardinalities are computed  at the end of the data stream.
Real-time applications such as anomaly detection require that user cardinalities are estimated on the fly.
To address these problems, we develop novel bit and register sharing algorithms,
which use a bit array and a register array to build a compact sketch of all users' connected items respectively.
Compared with previous bit and register sharing methods,
our algorithms exploit the dynamic properties of the bit and register arrays (e.g., the fraction of zero bits in the bit array at each time) to significantly improve the estimation accuracy,
and have low time complexity ($O(1)$) to update the estimations each time they observe a new user-item pair.
In addition, our algorithms are simple and easy to use, without  requirements to tune any parameter.
We evaluate the performance of our methods on real-world datasets.
The experimental results demonstrate that our methods are several times more accurate and faster than state-of-the-art methods using the same amount of memory.
\end{abstract}

\section{Introduction} \label{sec:introduction}
Many real-world networks are given in the form of graph streams.
Calling network is such an example with nodes representing users and an edge representing a call from one user to another.
When web surfing activities are modeled as a bipartite graph stream where users and items refer to network hosts and websites respectively,
an edge represents a visit by a user to a website.
Monitoring the cardinalities (or degrees) of users in these networks is fundamental for many applications such as network anomaly detection~\cite{Estan2003,Zhao2005,Nychis2008,YuNSDI2013},
where a user's cardinality is defined to be the number of \emph{\textbf{distinct}} users/items that the user connects to in the regular/bipartite graph stream of interest.
Due to the large-size and high-speed nature of these graph streams,
it is infeasible to collect the entire graph especially when the computation and memory resources are limited.
For example, network routers have fast but very small memories,
which leads their traffic monitoring modules  incapable to exactly compute the cardinalities of network users.
Therefore, it is important to develop fast and memory efficient algorithms to approximately compute user cardinalities over time.

Compared with using a counter to record a user's frequency (i.e., the number of times the user occurred) over time,
one needs to build a hash table of distinct occurred edges to handle edge duplicates in graph streams when computing user cardinalities.
Therefore, computing user cardinalities is more complex and difficult than computing user frequencies for large data streams,
and frequency estimation methods such as Count-Min sketch~\cite{Cormode2005improved} fails to approximate user cardinalities.
To address this challenge,
a variety of cardinality estimation methods such as Linear-Time Probabilistic Counting (LPC)~\cite{Whang1990} and HyperLogLog (HLL) \cite{FlajoletAOFA07} are developed to approximately compute cardinalities.
An LPC/HLL sketch consists of $m$ bits/registers, where $m$ is a parameter affecting the estimation accuracy.
Since user cardinalities are not known in advance and change over time,
one needs to set  $m$ large (e.g., thousand) to achieve reasonable accuracy for each user, whose  cardinality may vary over a large range.
However, this method wastes considerable memory because a large value of $m$ is not necessary for most users, which have small cardinalities.
To solve this problem, \cite{Zhao2005, Yoon2009, WangTIFS2012, XiaoSIGMETRICS2015} develop different virtual sketch methods to compress each user's LPC/HLL sketch into a large bit/register array shared by all users.
These virtual sketch methods build each user's virtual LPC/HLL sketch using $m$ bits/registers randomly selected from the large bit/register array.
This significantly reduces memory usage because each bit/register may be used by more than one user.
However,  bits/registers in a user's virtual LPC/HLL sketch may be contaminated by other users.
We refer to these bits/registers as ``\textbf{noisy}" bits/registers.
In practice, most users have small cardinalities and their virtual LPC/HLL sketches tend to contain many ``noisy" bits/registers,
which results in large estimation errors.
Another limitation of existing methods is that they are unable to report user cardinalities on the fly, because they are customized to estimate user cardinalities after all the data has been observed.
For  real-time applications like on-line anomaly detection, it is important to track user cardinalities in real-time.
For example, network monitoring systems are required to detect abnormal IP addresses such as super spreaders (i.e., IP addresses with cardinalities larger than a specified threshold) on the fly.
Moreover, online monitoring of IP address cardinalities over time also facilitates online detection of stealthy attacks launched from a subclass of IP addresses.

To address the above challenges, we develop two novel streaming algorithms FreeBS and FreeRS to accurately estimate user cardinalities over time.
We summarize our main contributions as:\\
\noindent$\bullet$ Compared with previous bit and register sharing methods, our algorithms FreeBS and FreeRS exploit the dynamic properties of the bit and register arrays over time (e.g., the fraction of zero bits in the bit array at each time) to significantly improve the estimation accuracy. To be more specific, FreeBS/FreeRS allows the number of bits/registers used by a user to dynamically increase as its cardinality increases over time and each user can use all shared bits/registers, which results in more accurate user cardinality estimations.

\noindent$\bullet$ Our algorithms report user cardinality estimations on the fly and allow to track user cardinalities in real-time. The time complexity is reduced from $O(m)$ in state-of-the-art methods CSE~\cite{Yoon2009} and vHLL~\cite{XiaoSIGMETRICS2015} to $O(1)$ for updating user cardinality estimations each time they observe a new user-item pair.

\noindent$\bullet$ We evaluate the performance of our methods on real-world datasets. Experimental results demonstrate that our methods are orders of magnitude faster and up to 10,000 times more accurate than state-of-the-art methods using the same amount of memory.

The rest of this paper is organized as follows.
The problem formulation is presented in Section~\ref{sec:problem}.
Section~\ref{sec:preliminaries} introduces preliminaries.
Section~\ref{sec:methods} presents our algorithms FreeBS and FreeRS.  
The performance evaluation and testing results are presented in Section~\ref{sec:results}.
Section~\ref{sec:related} summarizes related work.
Concluding remarks then follow.

\section{Problem Formulation} \label{sec:problem}
To formally define our problem,
we first introduce some notation.
Let $\Gamma=e^{(1)} e^{(2)} \cdots$ be the graph stream of interest consisting
of a sequence of edges.
Note that an edge in $\Gamma$ may appear more than once.
In this paper, we focus on bipartite graph streams consisting of edges between users and items.
Our methods however easily extend to regular graphs.
Let $S$ and $D$ denote the user and item sets, respectively.
For $t= 1, 2, \ldots$,
let $e^{(t)}=(s^{(t)}, d^{(t)})$ denote the $t^\text{th}$ edge of $\Gamma$,
where $s^{(t)}\in S$ and $d^{(t)}\in D$ are the user and the item of $e^{(t)}$ respectively.
Let $N_s^{(t)}$ denote the set of distinct items that user $s$ connects to before and including time $t$.
Define $n_s^{(t)}=|N_s^{(t)}|$ to be the cardinality of user $s$ at time $t$.
Then, $n^{(t)}=\sum_{s\in S}{|N_s^{(t)}|}$ is the sum of all user cardinalities at time $t$.
In this paper, we develop fast and accurate methods for estimating user cardinalities at times $t= 1, 2, \ldots$ using a limited amount of memory.
When no confusion arises,
we omit the superscript $(t)$ to ease exposition.

\section{Preliminaries}\label{sec:preliminaries}
\begin{figure*}[htb]
\center
\subfigure[CSE]{
\includegraphics[width=0.48\textwidth]{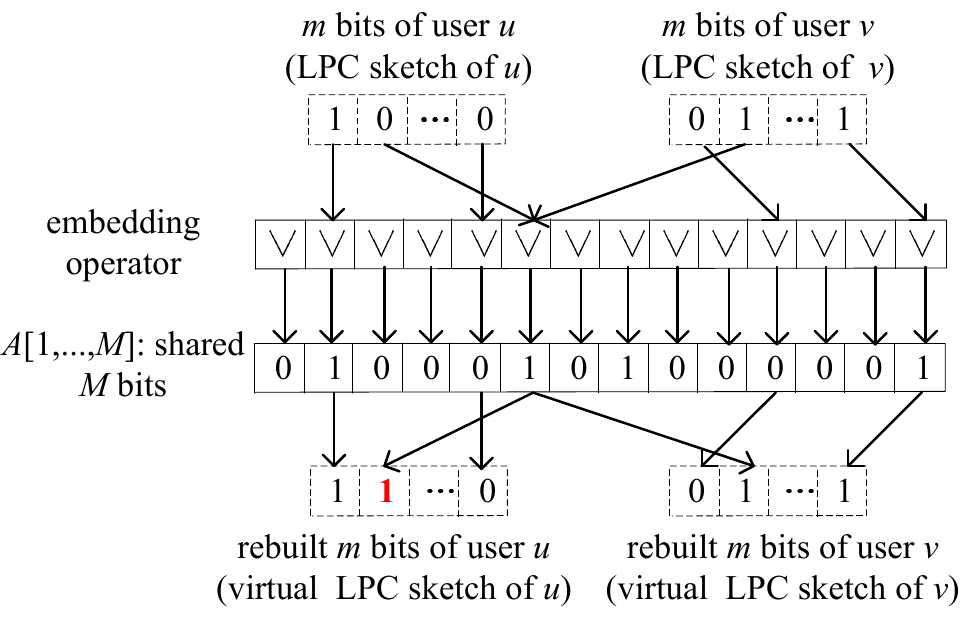}}
\subfigure[vHLL]{
\includegraphics[width=0.48\textwidth]{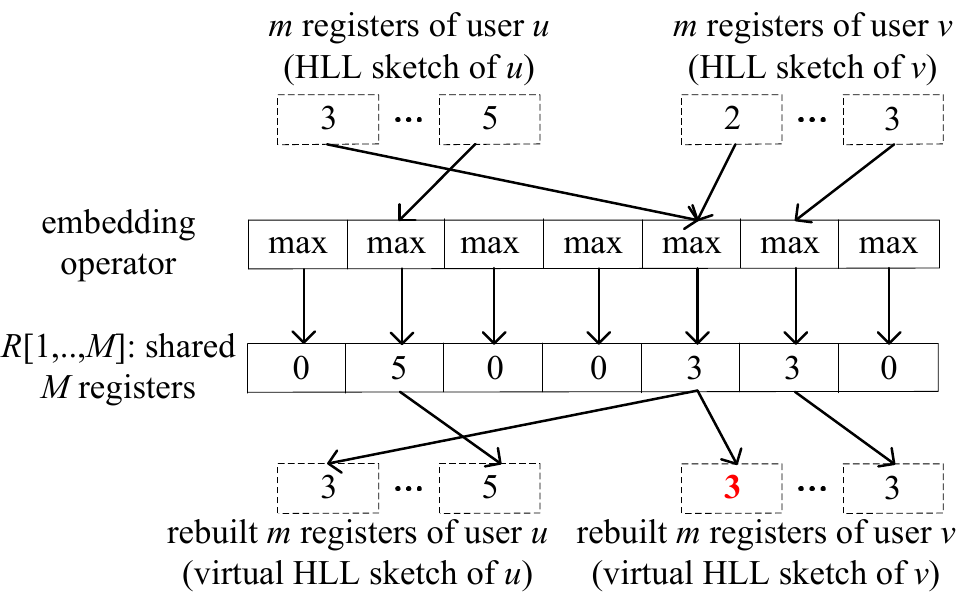}}
\caption{Overview of bit sharing method CSE and register sharing method vHLL. Virtual CSE/vHLL sketches of users may contain ``noisy" bits/registers (e.g., the bit and register in red and bold in the figure).}\label{fig:CSEandCHLL}
\end{figure*}
\subsection{Estimating a Single User's Cardinality}
\subsubsection{Linear-Time Probabilistic Counting}
For a user $s\in S$, Linear-Time Probabilistic Counting (LPC)~\cite{Whang1990} builds a sketch $B_s$ to store the set of items that $s$ connects to, i.e.,  $N_s^{(t)}$.
Formally, $B_s$ is defined as a bit array consisting of $m$ bits, which are initialized to zero.
Let $h(d)$ be a uniform hash function with range $\{1, \ldots, m\}$.
When user-item pair $(s,d)$ arrives, the $h(d)^\text{th}$ bit in $B_s$ is set to one, i.e., $B_s[h(d)] = 1$.
For any bit $B_s[i]$, $1\le i\le m$, the probability that it remains zero at time $t$ is
$P(B_s[i] = 0) = (1 - \frac{1}{m})^{n_s^{(t)}}$. Denote by $U_s^{(t)}$ the number of zero bits in $B_s$ at time $t$.
Then, the expectation of $U_s^{(t)}$ is computed as
$\mathbb{E}(U_s^{(t)}) = \sum_{i=1}^m P(B_s[i] = 0)\approx m e^\frac{-n_s^{(t)}}{m}$.
Based on the above equation, when $U_s^{(t)}>0$, Whang et al.~\cite{Whang1990} estimate $n_s^{(t)}$ as
\begin{equation*}
\hat n_s^{(t, \text{LPC})} = -m \ln \frac{U_s^{(t)}}{m}.
\end{equation*}
The range of $\hat n_s^{(t, \text{LPC})}$  is $[0, m\ln m]$,
and its expectation and variance are computed as
\[
\mathbb{E}(\hat n_s^{(t, \text{LPC})}) \approx n_s^{(t)} + \frac{1}{2}\left(e^{\frac{n_s^{(t)}}{m}} - \frac{n_s^{(t)}}{m} - 1\right),
\]
\[
{\rm Var}(\hat n_s^{(t, \text{LPC})}) \approx m \left(e^{\frac{n_s^{(t)}}{m}} - \frac{n_s^{(t)}}{m} - 1\right).
\]

\subsubsection{HyperLogLog}
To estimate the cardinality of user $s$,
HyperLogLog (HLL)~\cite{FlajoletAOFA07}
is developed based on the Flajolet-Martin (FM) sketch~\cite{Flajolet1985} $R_s$ consisting of $m$ registers $R_s[1], \ldots, R_s[m]$.
All $m$ registers are initialized to $0$.
For $1\le i\le m$, let $R_s^{(t)}[i]$ be the value of $R_s[i]$ at time $t$.
When a user-item pair $(s,d)$ arrives,
HLL maps the item into a pair of random variables $h(d)$ and $\rho(d)$, where $h(d)$ is an integer uniformly selected from \{1, \ldots, $m$\} at random
and $\rho(d)$ is drawn from a $Geometric(1/2)$ distribution,
$P(\rho(d)=k) = \frac{1}{2^k}$ for $k=1, 2, \ldots$.
\footnote{Functions $h(d)$ and $\rho(d)$ are usually implemented as: Let $\Phi(d)=\langle x_1x_2\cdots \rangle$ be the binary format of the output of a hash function $\Phi(d)$, and  $b=\lceil \log_2 m\rceil$. Then, $h(d)$ is defined as $h(d)=(x_1x_2\cdots x_b\mod m) + 1$ and $\rho(d)$ is defined as the number of leading zeros in $\langle x_{b+1} x_{b+2}\cdots \rangle$ plus one.}
Then, register $R_s[{h(d)}]$ is updated as
\[
R_s^{(t)}[h(d)]\gets \max\{R_s^{(t-1)}[h(d)], \rho(d)\}.
\]
At time $t$, Flajolet et al.~\cite{FlajoletAOFA07} estimate $n_s^{(t)}$ as
\[
\hat n_s^{(t, \text{HLL})}=\frac{\alpha_m m^2}{\sum_{i=1}^m 2^{-R^{(t)}[i]}},
\]
where $\alpha_m$ is the following constant to correct for bias
\[
\alpha_m =\left(m\int_0^{\infty} \left(\log_2 \frac{2+x}{1+x}\right)^m dx \right)^{-1}.
\]
The above formula for $\alpha_m$ is complicated.
In practice, $\alpha_m$ is computed numerically,
e.g., $\alpha_{16} \approx 0.673$, $\alpha_{32} \approx 0.697$, $\alpha_{64} \approx 0.709$, and $\alpha_m \approx 0.7213/(1 + 1.079/m)$ for $m \ge 128$.
The error of $\hat n_s^{(t, \text{HLL})}$ is analyzed as
\[
\lim_{n_s^{(t)}\rightarrow \infty}\frac{\mathbb{E}(\hat n_s^{(t, \text{HLL})})}{n_s^{(t)}} = 1 + \delta_1(n_s^{(t)}) + o(1),
\]
\[
\lim_{n_s^{(t)}\rightarrow \infty}\frac{\sqrt{{\rm Var}(\hat n_s^{(t, \text{HLL})})}}{n_s^{(t)}} = \frac{\beta_m}{\sqrt{m}} + \delta_2(n_s^{(t)}) + o(1).
\]
where $\delta_1$ and $\delta_2$ represent oscillating functions of a tiny amplitude (e.g., $|\delta_1(n_s^{(t)})|<5\times 10^{-5}$ and $|\delta_2(n_s^{(t)})|<5\times 10^{-4}$ as soon as $m\ge 16$) which can be safely neglected for all practical purposes, and $\beta_m$ is a constant for a specific $m$,
e.g., $\beta_{16} \approx 1.106$, $\beta_{32} \approx 1.070$, $\beta_{64} \approx 1.054$, $\beta_{128} \approx 1.046$, and $\beta_\infty \approx 1.039$.

Since $\hat n_s^{(t, \text{HLL})}$ is severely biased for small cardinalities,
HLL treats the HLL sketch $R_s$ as an LPC sketch (i.e., a bitmap of $m$ bits)  when $\frac{\alpha_m m^2}{\sum_{i=1}^m 2^{-R_s^{(t)}[i]}} < 2.5m$.
In this case, $n_s^{(t)}$ is estimated as $-m \ln \frac{\tilde U_s^{(t)}}{m}$,
where $\tilde U_s^{(t)}$ is the number of registers among $R_s[1]$, $\ldots$, $R_s[m]$ that equal 0 at time $t$.
Therefore, we easily find that LPC outperforms HLL for small cardinalities under the same memory usage.

\subsubsection{Discussions}
To compute all user cardinalities,
one can use an LPC or HLL sketch to estimate each user cardinality.
Clearly, using a small $m$, LPC and HLL will exhibit large errors for users with large cardinalities.
Most user cardinalities are small and assigning an LPC/HLL sketch with large $m$ to each user in order to accurately estimate large user cardinalities is wasteful as LPC and HLL do not require to set a large $m$ to achieve reasonable estimation accuracy for users with small cardinalities.
In practice, the user cardinalities are not known in advance and vary over time.
Therefore, it is difficult to set an optimal value of $m$ when using LPC and HLL to estimate all user cardinalities.
In the next subsection, we introduce state-of-the-art methods to address this problem,
and also discuss their shortcomings.
\subsection{Estimating All User Cardinalities}
\subsubsection{CSE: Compressing LPC Sketches of All Users into a Shared Bit Array}
As shown in Figure~\ref{fig:CSEandCHLL} (a), CSE~\cite{Yoon2009} consists of a large bit array $A$ and $m$ independent hash functions $f_1(s),...,f_m(s)$, each mapping users to $\{1, \ldots, M\}$, where $M$ is the length of the one dimensional bit array $A$.
Similar to LPC,
CSE builds a virtual LPC sketch for each user and embeds LPC sketches of all users in $A$.
For user $s$, its virtual LPC sketch $\hat B_s$ consists of $m$ bits selected randomly from $A$ by the group of hash functions $f_1(s),...,f_m(s)$, that is
\[
\hat B_s = (A[f_1(s)], \ldots, A[f_m(s)]).
\]
Each bit in $A$ is initially set to zero. When a user-item pair $(s,d)$ arrives, CSE sets the $h(d)^\text{th}$ bit in $\hat B_s$ to one.
Similar to LPC, $h(d)$ is a uniform hash function with range $\{1, \ldots, m\}$.
Since the $h(d)^\text{th}$ element in $\hat B_s$ is bit $A[f_{h(d)}(s)]$,
CSE only needs to set bit $A[f_{h(d)}(s)]$, i.e., $A[f_{h(d)}(s)]\gets 1$.
Let $\hat U_s^{(t)}$ be the number of zero bits in $\hat B_s$ and $U^{(t)}$ be the number of zero bits in $A$ at time $t$.
A user's virtual LPC sketch can be viewed as a regular LPC sketch containing ``noisy" bits (e.g., the bit in red and bold in Figure~\ref{fig:CSEandCHLL} (a)),
that are wrongly set from zero to one by items of other users.
To remove the estimation error introduced by ``noisy" bits,
Yoon et al.~\cite{Yoon2009} estimate $n_s^{(t)}$ as
\begin{equation*}
\hat n_s^{(t, \text{CSE})} = -m \ln \frac{\hat U_s^{(t)}}{m} + m\ln \frac{U^{(t)}}{M}.
\end{equation*}
On the right-hand side of the above equation,
the first term is the same as the regular LPC, and the second term corrects the error introduced by ``noisy" bits.
The bias and variance of $\hat n_s^{(t, \text{CSE})}$ are given by eqs. (23) and (24) in the original paper~\cite{Yoon2009}.

\subsubsection{vHLL: Compressing HLL Sketches of All Users into a Shared Bit Array}
Xiao et al.~\cite{XiaoSIGMETRICS2015} develop a register sharing method, vHLL, which extends the HLL method to estimate all user cardinalities.
vHLL consists of a list of $M$ registers $R[1], \ldots, R[M]$,
which are initialized to zero.
To maintain the virtual HLL sketch $\hat R_s$ of a user $s$, vHLL uses $m$ independent hash functions $f_1(s),...,f_m(s)$ to randomly select $m$ registers from all registers $R[1], \ldots, R[M]$,
where each function $f_1(s),...,f_m(s)$ maps users to $\{1, \ldots, M\}$.
Formally, $\hat R_s$ is defined as
\[
\hat R_s = (R[f_1(s)], \ldots, R[f_m(s)]).
\]

For $1\le i\le M$, let $R^{(t)}[i]$ be the value of $R[i]$ at time $t$.
When a user-item pair $(s,d)$ arrives,
it maps the item to a pair of two random variables $h(d)$ and $\rho(d)$,
where $h(d)$ is an integer uniformly selected from \{1, \ldots, $m$\} at random,
and $\rho(d)$ is a random integer drawn from a $Geometric(1/2)$ distribution,
which is similar to HLL.
We can easily find that the $h(d)^\text{th}$ element in the virtual HLL sketch of user $s$ is $R[f_{h(d)}(s)]$,
therefore, vHLL only needs to update register $R[f_{h(d)}(s)]$ as
\[
R^{(t)}[f_{h(d)}(s)]\gets \max\{R^{(t-1)}[f_{h(d)}(s)], \rho(d)\}.
\]
A user's virtual HLL sketch can be viewed as a regular HLL containing ``noisy" registers (e.g., the register in red and bold in Figure~\ref{fig:CSEandCHLL} (b)),
which are wrongly set by items of other users.
To remove the estimation error introduced by ``noisy" registers,
Xiao et al.~\cite{XiaoSIGMETRICS2015} estimate $n_s^{(t)}$ as
\begin{equation*}
\hat n_s^{(t, \text{vHLL})} = \frac{M}{M-m}\left(\frac{\alpha_m m^2}{\sum_{i=1}^m 2^{-R^{(t)}[f_i(s)]}} - \frac{m\alpha_M M}{\sum_{i=1}^M 2^{-R^{(t)}[i]}}\right),
\end{equation*}
where $\alpha_m$ is the same as that of HLL.
For the two terms between the parentheses on the right-hand side of the above equation,
the first term is the same as the regular HLL, and the second term corrects the error introduced by ``noisy" registers.
Similar to the regular HLL, the first term is replaced by $-m \ln \frac{\hat U_s^{(t)}}{m}$ when $\frac{\alpha_m m^2}{\sum_{i=1}^m 2^{-R^{(t)}[f_i(s)]}} < 2.5m$
where $\hat U_s^{(t)}$ is the number of registers among $R[f_1(s)]$, $\ldots$, $R[f_m(s)]$ that equal 0 at time $t$.
The expectation of $\hat n_s^{(t, \text{vHLL})}$ approximately equals $n_s^{(t)}$, that is,
$\mathbb{E}(\hat n_s^{(t, \text{vHLL})}) \approx n_s^{(t)}$.
The variance of $\hat n_s^{(t, \text{vHLL})}$ is approximately computed as
${\rm Var}(\hat n_s^{(t, \text{vHLL})}) \approx \frac{M^2}{(M-m)^2} (\frac{1.04^2}{m}(n_s^{(t)} + (n^{(t)} - n_s^{(t)})\frac{m}{M})^2
+(n^{(t)} - n_s^{(t)}) \frac{m}{M}\left(1-\frac{m}{M}\right) + \frac{(1.04n^{(t)}m)^2}{M^3})$,
where  $n^{(t)} = \sum_{s\in S} n_s^{(t)}$ counts  the total number of distinct user-item pairs occurred before and including time $t$.

\subsection{Unsolved Challenges}
\noindent\textbf{Challenge 1: It is difficult to set parameter $m$ for both CSE and vHLL.}
The estimation accuracy of CSE and vHLL highly depends on the value of $m$, as we can see in above discussions.
Increasing $m$ introduces more ``\emph{unused}" bits in virtual LPC sketches of occurred users, which can become contaminated with noise.
Here ``unused" bits refer to the bits in a user's virtual LPC sketch that no user-item pairs of the user are hashed into.
However, decreasing $m$ introduces large estimation errors for users with large cardinalities.
Similarly, vHLL also confronts the same challenge in determining an optimal $m$.
Later our experimental results will also verify that errors increase with $m$ for users with small cardinalities under CSE and vHLL.

\noindent\textbf{Challenge 2: It is computationally intensive to estimate user cardinalities for all values of $t$.}
At each time, both CSE and vHLL require time complexity $O(m)$ to compute the cardinality of a single user.
When applied to compute cardinalities for all users in $S$ at all times, CSE and vHLL have to be repeatedly called and will incur high computational cost, which
prohibits their application to high speed streams in an on-line manner.

\section{Our Methods}\label{sec:methods}
In this section, we present our streaming algorithms FreeBS and FreeRS for estimating user cardinalities over time.
FreeBS and FreeRS are designed based on two novel bit sharing and register sharing techniques, respectively.
The basic idea behind our methods can be summarized as:
Unlike vHLL/CSE mapping each user's items into $m\ll M$ bits/registers,
FreeBS/FreeRS randomly maps user-item pairs into all $M$ bits/registers in the bit/register array.
Thus, users with larger cardinalities (i.e., connecting to a larger number of items) tend to use more bits/registers.
For each user-item pair $e^{(t)}=(s^{(t)}, d^{(t)})$ occurred at time $t$,
we discard it when updating $e^{(t)}$ does not change any bit/register in the bit/register array shared by all users.
Otherwise, $e^{(t)}$ is a new user-item pair that does not occur before time $t$,
and we increase the cardinality estimation of user $s^{(t)}$ by $\frac{1}{q^{(t)}}$,
where $q^{(t)}$ is defined as the probability that a new user-item pair changes any bit/register in the bit/register array at time $t$.
To some extent, the above procedure can be viewed as a sampling method
such that each new user-item pair arriving at time $t$ is sampled with probability $q^{(t)}$
and each user's cardinality is estimated using the Horvitz-Thompson estimator~\cite{horvitz1952gsw}.

\subsection{FreeBS: Parameter-Free Bit Sharing}
\noindent\textbf{Data Structure.}
The pseudo-code for FreeBS is shown as Algorithm~\ref{alg:FreeBS}.
FreeBS consists of a one-dimensional bit array $B$ of length $M$,
where each bit $B[j]$, $1\le j\le M$, is initialized to zero.
In addition, FreeBS uses a hash function $h^*(e)$ to uniformly and independently map each user-item pair $e=(s,d)$ to an integer in $\{1, \ldots, M\}$ at random,
i.e., $P(h^*(e)=i)=\frac{1}{M}$, and $P(h^*(e)=i\wedge h^*(e')=i'|e\ne e')=\frac{1}{M^2}$, $i, i'\in \{1, \ldots, M\}$.
Note that $h^*(e)$ differs from the hash function $h(s)$ used by CSE that maps \textbf{ user $s$} to an integer in $\{1, \ldots, M\}$ at random.

\begin{algorithm}
\SetKwFunction{insert}{insert}
\SetKwFunction{delete}{delete}
\SetKwFunction{continue}{continue}
\SetKwFunction{MaxRankEdge}{MaxRankEdge}
\SetKwFunction{ComputeEdgeLabel}{ComputeEdgeLabel}
\SetKwFunction{UpdateEdgeLabel}{UpdateEdgeLabel}
\SetKwFunction{UpdateTriCounts}{UpdateTriCounts}
\SetKwInOut{Input}{input}
\SetKwInOut{Output}{output}
\BlankLine
$B[1, \ldots, M]\gets [0, \ldots, 0]$\;
$\hat n_s^\text{(FreeBS)} \gets 0$, $s\in S$\;
$m_0\gets M$\;
\ForEach {$e=(s, d)$ in $\Gamma$}{
    \If {$B[h^*(e)]==0$}{
        $B[h^*(e)]\gets 1$\;
        $\hat n_s^\text{(FreeBS)}\gets \hat n_s^\text{(FreeBS)} + \frac{M}{m_0}$\;
        $m_0\gets m_0 - 1$\;
    }
}
\caption{The pseudo-code for FreeBS.}\label{alg:FreeBS}
\end{algorithm}

\noindent\textbf{Update Procedure.}
When a user-item pair $e=(s,d)$ arrives at time $t$,
FreeBS first computes a random variable $h^*(e)$,
and then sets $B[h^*(e)]$ to one, i.e., $B[h^*(e)]\gets 1.$
Let $B_0^{(t)}$ denote the set of the indices corresponding to zero bits in $B$ at time $t$.
Formally, $B_0^{(t)}$ is defined as $B_0^{(t)}=\{i: B^{(t)}[i]=0, 1\le i\le M\}.$
Let $\hat n_s^{(t, \text{FreeBS})}$ denote the cardinality estimate for user $s$ at time $t$.
We initialize $\hat n_s^{(0, \text{FreeBS})}$ to 0.
Next, we describe how $\hat n_s^{(t, \text{FreeBS})}$ is computed on-line.
Let $m_0^{(t)} = |B_0^{(t)}|$ denote the number of zero bits in $B$ at time $t$.
Let $q_\text{B}^{(t)}$ denote the probability of $e$ changing a bit in $B$ from 0 to 1 at time $t$.
Formally, $q_\text{B}^{(t)}$ is defined as
\begin{equation*}
q_\text{B}^{(t)} = \sum_{i\in B_0^{(t-1)}}  P(h^*(e)=i)=  \frac{|B_0^{(t-1)}|}{M} = \frac{m_0^{(t-1)}}{M}.
\end{equation*}
Let $\mathbf{1}(\mathbb{P})$ denote the indicator function that equals 1 when predicate $\mathbb{P}$ is true and 0 otherwise.
Besides setting $B[h^*(e)]$ to 1 at time $t$ with the arrival of user-item pair $e=(s,d)$,
we also update the cardinality estimate of user $s$ as
\[
\hat n_s^{(t, \text{FreeBS})}\gets \hat n_s^{(t-1, \text{FreeBS})} + \frac{\mathbf{1}(B^{(t-1)}[h^*(e)] = 0)}{q_\text{B}^{(t)}}.
\]
For any other user $s'\in S\setminus\{s\}$, we keep its cardinality estimate unchanged, i.e., $\hat n_{s'}^{(t, \text{FreeBS})}\gets \hat n_{s'}^{(t-1, \text{FreeBS})}$.
We easily find that $q_\text{B}^{(t)}$ can be fast computed incrementally.
That is, we initialize $q_\text{B}^{(1)}$ to 1, and incrementally compute $q_\text{B}^{(t+1)}$ as
\begin{equation*}
q_\text{B}^{(t+1)} \gets q_\text{B}^{(t)}-\frac{\mathbf{1}(B^{(t-1)}[h^*(e)] = 0)}{M}, \quad t\ge 1.
\end{equation*}
Hence,  the time complexity of FreeBS for processing each user-item pair is $O(1)$.

\noindent\textbf{Error Analysis.}
Let $T_s^{(t)}$ denote the set of the first occurrence times of user-item pairs associated with user $s$ in stream $\Gamma$.
Formally, we define $T_s^{(t)}$ as
\[
T_s^{(t)} = \{i: s^{(i)}=s \wedge e^{(j)}\ne e^{(i)},  0<j< i\le t\}.
\]

\begin{theorem}\label{theorem:FreeBS}
The expectation and variance of $\hat n_s^{(t, \text{FreeBS})}$ are
\[
\mathbb{E}(\hat n_s^{(t, \text{FreeBS})}) = n_s^{(t)},
\]
\[
\text{Var}(\hat n_s^{(t, \text{FreeBS})}) = \sum_{i\in T_s^{(t)}} \mathbb{E}(\frac{1}{q_\text{B}^{(i)}}) - n_s^{(t)}\le n_s^{(t)}\left(\mathbb{E}(\frac{1}{q_\text{B}^{(t)}}) - 1\right),
\]
where
\begin{equation*}
\begin{split}
\mathbb{E}(\frac{1}{q_\text{B}^{(i)}}) &= \sum_{j = 1}^M \frac{M\binom{n^{(i)}}{j} \sum_{k=0}^{j-1} (-1)^k \binom{j}{k} (\frac{j-k}{M})^{n^{(i)}}}{M-j}\\
&\approx e^{\frac{n^{(i)}}{M}} \left(1+\frac{1}{M}\left(e^{\frac{n^{(i)}}{M}} - \frac{n^{(i)}}{M} - 1\right)\right).
\end{split}
\end{equation*}
\end{theorem}
\begin{pf}
Let $\delta_{e}$ denote an indicator that equals 1 when updating a user-item pair $e$ incurs a value change of $B[h^*(e)]$ (i.e., $B[h^*(e)]$ changes from 0 to 1),
and 0 otherwise.
We easily have 
\[
\hat n_s^{(t, \text{FreeBS})} = \sum_{i\in T_s^{(t)}} \frac{\delta_{e^{(i)}}}{q_\text{B}^{(i)}}.
\]
For each $\delta_{e^{(i)}}$, we have
\[
\mathbb{E}(\delta_{e^{(i)}}|q_\text{B}^{(i)}) = q_\text{B}^{(i)}, \quad 1\le i\le t,
\]
\[
\text{Var}(\delta_{e^{(i)}}|q_\text{B}^{(i)}) = \mathbb{E}(\delta_{e^{(i)}}^2|q_\text{B}^{(i)}) - (\mathbb{E}(\delta_{e^{(i)}}|q_\text{B}^{(i)}))^2 = q_\text{B}^{(i)} - (q_\text{B}^{(i)})^2.
\]
Define $Q_\text{B}^{(t)}=\{q_\text{B}^{(1)}, \ldots, q_\text{B}^{(t)}\}.$
Then, we have
\begin{equation*}
\begin{split}
\mathbb{E}(\hat n_s^{(t, \text{FreeBS})}| Q_\text{B}^{(t)}) &= \mathbb{E}\left(\sum_{i\in T_s^{(t)}} \frac{\delta_{e^{(i)}}}{q_\text{B}^{(i)}}\middle| Q_\text{B}^{(t)}\right)\\
&= \sum_{i\in T_s^{(t)}} \frac{\mathbb{E}(\delta_{e^{(i)}}|Q_\text{B}^{(t)})}{q_\text{B}^{(i)}}\\
&= \sum_{i\in T_s^{(t)}} \frac{P(\delta_{e^{(i)}}=1|Q_\text{B}^{(t)})}{q_\text{B}^{(i)}}\\
&= n_s^{(t)}.
\end{split}
\end{equation*}
Given $Q_\text{B}^{(t)}$,  random variables $\delta_{e^{(i)}}$, $i\in T_s^{(t)}$, are independent of each other.
Then, we have
\[
\mathbb{E}(\hat n_s^{(t, \text{FreeBS})}) = \mathbb{E}(\mathbb{E}(\hat n_s^{(t, \text{FreeBS})}|Q_\text{B}^{(t)})) = \mathbb{E}(n_s^{(t)}) = n_s^{(t)}.
\]
The variance of $\hat n_s^{(t, \text{FreeBS})}$ is computed as
\begin{equation*}
\begin{split}
\text{Var}(\hat n_s^{(t, \text{FreeBS})}|Q_\text{B}^{(t)}) &= \text{Var}\left(\sum_{i\in T_s^{(t)}} \frac{\delta_{e^{(i)}}}{q_\text{B}^{(i)}}\middle|Q_\text{B}^{(t)}\right)\\
&= \sum_{i\in T_s^{(t)}} \frac{\text{Var}(\delta_{e^{(i)}}|Q_\text{B}^{(t)})}{(q_\text{B}^{(i)})^2}\\
&= \sum_{i\in T_s^{(t)}} \frac{q_\text{B}^{(i)} - (q_\text{B}^{(i)})^2}{(q_\text{B}^{(i)})^2}\\
&= \sum_{i\in T_s^{(t)}} \frac{1}{q_\text{B}^{(i)}} - n_s^{(t)}.
\end{split}
\end{equation*}
Since $\text{Var}(\mathbb{E}(\hat n_s^{(t, \text{FreeBS})}|Q_\text{B}^{(t)}))=0$,
using the equation $\text{Var}(X) = \text{Var}(\mathbb{E}(X|Y)) + \mathbb{E}(\text{Var}(X|Y))$,
  we have
\begin{equation*}
\begin{split}
\text{Var}(\hat n_s^{(t, \text{FreeBS})}) &= \mathbb{E}(\text{Var}(\hat n_s^{(t, \text{FreeBS})}|Q_\text{B}^{(t)}))\\
&=\mathbb{E}(\sum_{i\in T_s^{(t)}} \frac{1}{q_\text{B}^{(i)}}) - n_s^{(t)}\\
&=\sum_{i\in T_s^{(t)}} \mathbb{E}(\frac{1}{q_\text{B}^{(i)}}) - n_s^{(t)}.
\end{split}
\end{equation*}

In what follows we derive the formula for $\mathbb{E}(\frac{1}{q_\text{B}^{(i)}})$.
For $j$ specific distinct bits in $B$, there exist $j!\tau(n^{(i)}, j)$ ways to map $n^{(i)}$ distinct user-item pairs occurred in stream $\Gamma$ before and including time $i$ into these bits given that each bit has at least one user-item pair,
where $\tau(n^{(i)}, j)$, the Stirling number of the second kind~\cite{Abramowitz1964}, is computed as
\[
\tau(n^{(i)}, j)=\sum_{k=0}^{j-1} (-1)^k \binom{j}{k} (j-k)^{n^{(i)}},~0< j \le n^{(i)}.
\]
In addition, there exist $\binom{M}{j}$
ways to select $j$ distinct bits from $B$, therefore we have
\[
P(m_0^{(i)} = M - j|n^{(i)}) = \frac{\binom{M}{j} j!\tau(n^{(i)}, j)}{M^{n^{(i)}}}.
\]
Then, we   have
\begin{equation*}
\mathbb{E}(\frac{1}{q_\text{B}^{(i)}}) = \sum_{j = 1}^M \frac{M\binom{n^{(i)}}{j} \sum_{k=0}^{j-1} (-1)^k \binom{j}{k} (\frac{j-k}{M})^{n^{(i)}}}{M-j}.
\end{equation*}
Next, we introduce a method to approximately compute $\mathbb{E}(\frac{1}{q_\text{B}^{(i)}})$.
We expand the function $\frac{1}{q_\text{B}^{(i)}}$ by its Taylor series around $\mathbb{E}(q_\text{B}^{(i)})$ as
\begin{equation*}
\begin{split}
\mathbb{E}(\frac{1}{q_\text{B}^{(i)}}) &\approx  \mathbb{E}\left(\frac{1}{\mathbb{E}(q_\text{B}^{(i)})} - \frac{ q_\text{B}^{(i)}-\mathbb{E}(q_\text{B}^{(i)})}{(\mathbb{E}(q_\text{B}^{(i)}))^2}
+  \frac{ (q_\text{B}^{(i)}-\mathbb{E}(q_\text{B}^{(i)}))^2}{(\mathbb{E}(q_\text{B}^{(i)}))^3}\right)\\
&=\frac{1}{\mathbb{E}(q_\text{B}^{(i)})} + \frac{\text{Var}(q_\text{B}^{(i)})}{(\mathbb{E}(q_\text{B}^{(i)}))^3}.
\end{split}
\end{equation*}
From~\cite{Whang1990} (eqs.(5) and (6) in~\cite{Whang1990}),
we easily have
$\mathbb{E}(q_\text{B}^{(i)}) = e^{-\frac{n^{(i)}}{M}}$
and $\text{Var}(q_\text{B}^{(i)}) =\frac{1}{M} e^{-\frac{n^{(i)}}{M}}(1-(1+\frac{n^{(i)}}{M})e^{-\frac{n^{(i)}}{M}})$.
Then, we   obtain
$\mathbb{E}(q_\text{B}^{(i)}) \approx e^{\frac{n^{(i)}}{M}} (1+\frac{1}{M}(e^{\frac{n^{(i)}}{M}} - \frac{n^{(i)}}{M} - 1))$.
\end{pf}

\subsection{FreeRS: Parameter-Free Register Sharing}
\noindent\textbf{Data Structure.}
The pseudo-code for FreeRS is shown as Algorithm~\ref{alg:FreeRS}.
FreeRS consists of $M$ registers $R[1]$, $\ldots$, $R[M]$, which are initialized to zero.
In addition, FreeRS also uses a hash function $h^*(e)$ to randomly map each user-item pair $e=(s,d)$ to an integer in $\{1, \ldots, M\}$
and another function $\rho^*(e)$ that maps $e$ to a random integer in $\{1, 2, \ldots\}$ according to a $Geometric(1/2)$ distribution.
Note that $h^*(e)$ and $\rho^*(e)$ differ from hash functions $h(s)$ and $\rho(s)$ used by vHLL, which map \textbf{user $u$} to a random integer in $\{1, \ldots, M\}$ and $\{1, 2, \ldots\}$, respectively.

\noindent\textbf{Update Procedure.}
When user-item pair $e=(s,d)$ arrives at time $t$,
FreeRS first computes two random variables $h^*(e)$ and $\rho^*(e)$,
and then updates $R^{(t)}[h^*(e)]$ as
\[
R^{(t)}[h^*(e)]\gets \max\{R^{(t-1)}[h^*(e)], \rho^*(e)\}.
\]
Let $q_\text{R}^{(t)}$ denote the probability of $e$ changing the value of a register among $R[1], \ldots, R[M]$ at time $t$.
Formally, $q_\text{R}^{(t)}$ is defined as
\begin{equation*}
\begin{split}
q_\text{R}^{(t)} &= \sum_{j=1}^M  P(h^*(e)=j\wedge R^{(t)}[j] > R^{(t-1)}[j])\\
&= \frac{\sum_{j=1}^M 2^{-R^{(t-1)}[j]}}{M}.
\end{split}
\end{equation*}

\begin{algorithm}
\SetKwFunction{insert}{insert}
\SetKwFunction{delete}{delete}
\SetKwFunction{continue}{continue}
\SetKwFunction{MaxRankEdge}{MaxRankEdge}
\SetKwFunction{ComputeEdgeLabel}{ComputeEdgeLabel}
\SetKwFunction{UpdateEdgeLabel}{UpdateEdgeLabel}
\SetKwFunction{UpdateTriCounts}{UpdateTriCounts}
\SetKwInOut{Input}{input}
\SetKwInOut{Output}{output}
\BlankLine
$R[1, \ldots, M]\gets [0, \ldots, 0]$; $q_\text{R}\gets 1$\;
$\hat n_s^\text{(FreeRS)} \gets 0$, $s\in S$\;

\ForEach {$e=(s, d)\in \Gamma$}{
    \If {$\rho^*(e) > R[h^*(e)]$}{
        $q_\text{R}\gets q_\text{R} + \frac{2^{-\rho^*(e)} - 2^{-R[h^*(e)]}}{M}$\;
        $R[h^*(e)]\gets \rho^*(e)$\;
        $\hat n_s^\text{(FreeRS)}\gets \hat n_s^\text{(FreeRS)} + \frac{1}{q_\text{R}}$\;
    }
}
\caption{The pseudo-code for FreeRS.}\label{alg:FreeRS}
\end{algorithm}

Let $\hat n_s^{(t, \text{FreeRS})}$ denote the cardinality estimate of user $s$ at time $t$.
When user-item pair $e=(s,d)$ arrives at time $t$,
we update the cardinality estimate of user $s$ as
\[
\hat n_s^{(t, \text{FreeRS})}\gets \hat n_s^{(t-1, \text{FreeRS})} + \frac{\mathbf{1}(R^{(t)}[h^*(e)]\ne R^{(t-1)}[h^*(e)])}{q_\text{R}^{(t)}}.
\]
For any other user $s'\in S\setminus\{s\}$, we keep its cardinality estimate unchanged, i.e., $\hat n_{s'}^{(t, \text{FreeRS})}\gets \hat n_{s'}^{(t-1, \text{FreeRS})}$.
Similar to $q_\text{B}^{(t)}$,
we compute $q_\text{R}^{(t)}$ incrementally.
In detail,
we initialize $q_\text{R}^{(1)} = 1$ and incrementally compute $q_\text{R}^{(t+1)}$ as
\begin{equation*}
\begin{split}
q_\text{R}^{(t+1)} &\gets q_\text{R}^{(t)}+\\
&\frac{2^{-\rho^*(e)} - 2^{-R[h^*(e)]}}{m} \mathbf{1}(R^{(t)}[h^*(e)]\ne R^{(t-1)}[h^*(e)]).
\end{split}
\end{equation*}
Hence, the time complexity of FreeRS for processing each user-item pair is also $O(1)$.

\noindent\textbf{Error Analysis.}
We derive the error of $\hat n_s^{(t, \text{FreeRS})}$ as follows:
\begin{theorem}\label{theorem:FreeRS}
The expectation and variance of $\hat n_s^{(t, \text{FreeRS})}$ are
\[
\mathbb{E}(\hat n_s^{(t, \text{FreeRS})}) = n_s^{(t)},
\]
\[
\text{Var}(\hat n_s^{(t, \text{FreeRS})}) =  \sum_{i\in T_s^{(t)}}\mathbb{E}(\frac{1}{q_\text{R}^{(i)}}) - n_s^{(t)}\le n_s^{(t)}\left(\mathbb{E}(\frac{1}{q_\text{R}^{(t)}}) - 1\right),
\]
where
\begin{equation*}
\begin{split}
\mathbb{E}(\frac{1}{q_\text{R}^{(i)}}) &= \sum_{k_1,\ldots,k_M\ge 0} \frac{\sum_{n_1+\ldots+n_M=n^{(i)}}\binom{n^{(i)}}{n_1,\ldots,n_M}\Pi_{j=1}^M \gamma_{n_j, k_j}}{M^{n^{(i)}-1}\sum_{j=1}^M 2^{-k_j}},\\
\end{split}
\end{equation*}
with
\[
\gamma_{n_j, k_j}=\left\{
                    \begin{array}{ll}
                      (1-2^{-k_j})^{n_j} - (1-2^{-k_j+1})^{n_j}, & n_j>0, k_j>0 \\
                      0, & n_j>0, k_j = 0 \\
                      1, & n_j = 0, k_j = 0.
                    \end{array}
                  \right.
\]
$\mathbb{E}(\frac{1}{q_\text{R}^{(i)}})$ is approximately $\frac{1.386 n^{(i)}}{M}$ when $n^{(i)}>2.5 M$.
\end{theorem}
\begin{pf}
Let $\xi_{e}$ denote an indicator equal to 1 when processing a user-item pair $e$ incurs a value change of any $R[1], \ldots, R[M]$,
and 0 otherwise.
We have
\[
\hat n_s^{(t, \text{FreeRS})} = \sum_{i\in T_s^{(t)}} \frac{\xi_{e^{(i)}}}{q_\text{R}^{(i)}}.
\]
For each $\xi_{e^{(i)}}$, we have
\[
\mathbb{E}(\xi_{e^{(i)}}|q_\text{R}^{(i)}) = q_\text{R}^{(i)}, \quad 1\le i\le t,
\]
\[
\text{Var}(\xi_{e^{(i)}}|q_\text{R}^{(i)}) = \mathbb{E}(\xi_{e^{(i)}}^2|q_\text{R}^{(i)}) - (\mathbb{E}(\xi_{e^{(i)}}|q_\text{R}^{(i)}))^2 = q_\text{R}^{(i)} - (q_\text{R}^{(i)})^2.
\]
Similar to FreeBS, define
\[
Q_\text{R}^{(t)}=\{q_\text{R}^{(1)}, \ldots, q_\text{R}^{(t)}\}.
\]
Then, we have
\begin{equation*}
\mathbb{E}(\hat n_s^{(t, \text{FreeRS})}|Q_\text{R}^{(t)}) = \mathbb{E}(\sum_{i\in T_s^{(t)}} \frac{\xi_{e^{(i)}}}{q_\text{R}^{(i)}} \Bigm| Q_\text{R}^{(t)})= n_s^{(t)}.
\end{equation*}
Therefore, we obtain
\[
\mathbb{E}(\hat n_s^{(t, \text{FreeRS})}) = \mathbb{E}(\mathbb{E}(\hat n_s^{(t, \text{FreeRS})}|Q_\text{R}^{(t)})) = \mathbb{E}(n_s^{(t)}) = n_s^{(t)}.
\]
Given $Q_\text{R}^{(t)}$, all $\xi_{e^{(i)}}$ ($i\in T_s^{(t)}$) are independent of each other.
Similar to FreeBS, the variance of $\hat n_s^{(t, \text{FreeRS})}$ given $Q_\text{R}^{(t)}$ is
\begin{equation*}
\text{Var}(\hat n_s^{(t, \text{FreeRS})}|Q_\text{R}^{(t)}) = \sum_{i\in T_s^{(t)}} \frac{1}{q_\text{R}^{(i)}} - n_s^{(t)}.
\end{equation*}
It is easily shown that $\text{Var}(\mathbb{E}(\hat n_s^{(t, \text{FreeRS})}|Q_\text{R}^{(t)}))=0$.
Using equation $\text{Var}(X) = \text{Var}(\mathbb{E}(X|Y)) + \mathbb{E}(\text{Var}(X|Y))$,
we have
\begin{equation*}
\begin{split}
\text{Var}(\hat n_s^{(t, \text{FreeRS})}) &= \mathbb{E}(\text{Var}(\hat n_s^{(t, \text{FreeRS})}|Q_\text{R}^{(t)}))\\
=&\mathbb{E}(\sum_{i\in T_s^{(t)}} \frac{1}{q_\text{R}^{(i)}}) - n_s^{(t)}=\sum_{i\in T_s^{(t)}} \mathbb{E}(\frac{1}{q_\text{R}^{(i)}}) - n_s^{(t)}.
\end{split}
\end{equation*}

In what follows we derive the formula for $\mathbb{E}(\frac{1}{q_\text{R}^{(i)}})$.
Using $h^*(\cdot)$, FreeRS randomly splits stream $\Gamma$ into $M$ sub-streams $\Gamma_j$, $1\le j\le M$.
Each $R[j]$ tracks the maximum value of function $\rho^*(\cdot)$ for user-item pairs in sub-stream $\Gamma_j$.
At time $i$, assume that $n_j$ distinct user-item pairs have occurred in $\Gamma_j$.
Then, $P(R^{(i)}[j] = k_j|n_j) = \gamma_{n_j, k_j}$.
Therefore,
\begin{equation*}
\begin{split}
&P(R^{(i)}[1] = k_1, \ldots, R^{(t)}[M] = k_M|n^{(i)})=\\
&\frac{\sum_{n_1+\ldots+n_M=n^{(i)}}\binom{n^{(i)}}{n_1,\ldots,n_M}\Pi_{j=1}^M \gamma_{n_j, k_j}}{M^{n^{(i)}}}.\\
\end{split}
\end{equation*}
An exact expression for $\mathbb{E}(\frac{1}{q_\text{R}^{(i)}})$ is easily derived.
However, it is too complex to analyze.
Hence, we introduce a method to approximate $\mathbb{E}(\frac{1}{q_\text{R}^{(i)}})$.
From~\cite{FlajoletAOFA07},
we have
$\mathbb{E}(\frac{\alpha_M M}{q_\text{R}^{(i)}}) = \alpha_M M \mathbb{E}(\frac{1}{q_\text{R}^{(i)}}) \approx n^{(i)}$ for $n^{(i)}>2.5 M$.
Therefore, $\mathbb{E}(\frac{1}{q_\text{R}^{(i)}})\approx \frac{n^{(i)}}{\alpha_M M}\approx \frac{1.386 n^{(i)}}{M}$.
\end{pf}
\subsection{Discussions} \label{sec:discussion}
\noindent\textbf{FreeBS vs CSE.} FreeBS outperforms CSE in three aspects: (1) FreeBS can estimate cardinalities up to $\sum_{i=1}^M \frac{M}{i} \approx  M\ln M$, which is larger than the maximum cardinality $m\ln m$ allowed by CSE;
(2) FreeBS exhibits a smaller estimation error than CSE.
From~\cite{TaoWGH17}, we find that the expectation and the variance of estimation $\hat n_s^{(t, \text{CSE})}$ given by CSE are
\[
\mathbb{E}(\hat n_s^{(t, \text{CSE})}) \approx n_s^{(t)} + \frac{1}{2}\left(\mathbb{E}(\frac{1}{q^{(t)}}) e^{\frac{n_s^{(t)}}{m}} - \frac{n_s^{(t)}}{m} - 1\right),
\]
\[
{\rm Var}(\hat n_s^{(t, \text{CSE})}) \approx m \left(\mathbb{E}(\frac{1}{q^{(t)}}) e^{\frac{n_s^{(t)}}{m}} - \frac{n_s^{(t)}}{m} - 1\right),
\]
where $q^{(t)} =\frac{U^{(t)}}{M}$ is the fraction of zero bits in the shared bit array at time $t$.
We find that CSE exhibits a large bias when $n_s^{(t)}\gg m$,
while FreeBS is unbiased.
When $m$ is large (e.g., $m$ approaches to $M$), FreeBS and CSE perform nearly the same bit setting operations but use their different cardinality estimators.
Then, we have $\mathbb{E}(\frac{1}{q^{(t)}}) \approx \mathbb{E}(\frac{1}{q_\text{B}^{(t)}})$.
From Theorem~\ref{theorem:FreeRS}, therefore, we easily have
\begin{equation*}
\begin{split}
{\rm Var}(\hat n_s^{(t, \text{CSE})})&\gtrapprox m \left(\mathbb{E}(\frac{1}{q^{(t)}}) (1+\frac{n_s^{(t)}}{m}) - \frac{n_s^{(t)}}{m} - 1\right)\\
 &>n_s^{(t)} \mathbb{E}(\frac{1}{q^{(t)}}) - n_s^{(t)}\gtrapprox{\rm Var}(\hat n_s^{(t, \text{FreeBS})});\
\end{split}
\end{equation*}
3) FreeBS has complexity $O(1)$ to update user cardinality estimates each time it observes a new user-item pair,
while CSE has complexity $O(m)$.

\noindent\textbf{FreeRS vs vHLL.} FreeRS and vHLL both can estimate cardinalities up to $2^{2^w}$,
where $w$ is the number of bits in a register.
However, FreeRS outperforms vHLL in two aspects:
1) FreeRS has complexity $O(1)$ to update user cardinality estimates  each time it observes a new user-item pair,
while vHLL has complexity $O(m)$;
2) FreeRS exhibits a smaller estimation error in comparison with vHLL. From Theorem~\ref{theorem:FreeRS}, then we have $\text{Var}(\hat n_s^{(t, \text{FreeRS})})\le n_s^{(t)}(\mathbb{E}(\frac{1}{q_\text{R}^{(t)}}) - 1)\approx n_s^{(t)}(\frac{n^{(t)}}{M\alpha_M} - 1)<\frac{1.386 n^{(t)} n_s^{(t)}}{M}$, while the variance of vHLL is
\begin{equation*}
\begin{split}
{\rm Var}(\hat n_s^{(t, \text{vHLL})}) &\gtrapprox(\frac{M}{M-m})^2 \times \frac{1.04^2}{m} \times  2n^{(t)} n_s^{(t)} \frac{m}{M}(1-\frac{m}{M})\\
&=\frac{2.163 n^{(t)} n_s^{(t)}}{M-m}>\frac{2.163 n^{(t)} n_s^{(t)}}{M}.
\end{split}
\end{equation*}

\noindent\textbf{FreeBS vs FreeRS.}
We observe that
1) FreeBS is faster than FreeRS. For each coming user-item pair $e$, FreeBS only computes $h^*(e)$ to select and set a bit, but FreeRS needs to compute both $h^*(e)$ and $\rho^*(e)$ to select and update a register;
2) Under the same memory usage, we compare the accuracy of FreeBS using $M$ bits and FreeRS using $M/w$ registers,
where $w$ is the number of bits in a register.
From Theorems~\ref{theorem:FreeBS} and ~\ref{theorem:FreeRS}, we have
\[
\text{Var}(\hat n_s^{(t, \text{FreeBS})}) =  \sum_{i\in T_s^{(t)}}\mathbb{E}(\frac{1}{q_\text{B}^{(i)}}) - n_s^{(t)},
\]
\[
\text{Var}(\hat n_s^{(t, \text{FreeRS})}) =  \sum_{i\in T_s^{(t)}}\mathbb{E}(\frac{1}{q_\text{R}^{(i)}}) - n_s^{(t)},
\]
where $\mathbb{E}(\frac{1}{q_\text{B}^{(i)}})\approx e^{\frac{n^{(i)}}{M}}$ and $\mathbb{E}(\frac{1}{q_\text{R}^{(i)}})\approx \frac{1.386 w n^{(i)}}{M} <  e^{\frac{n^{(i)}}{M}}$ when $\frac{n^{(i)}}{M}\ge 0.772 w$.
Therefore, FreeRS is more accurate than FreeBS for users not appearing among the first $0.772 w M$ distinct user-item pairs presented on stream $\Gamma$.
Flajolet et al.~\cite{FlajoletAOFA07} observe that HLL exhibits a large error for estimating small cardinalities.
To solve this problem, they view a register of HLL as a bit of LPC and estimate the cardinality based on the fraction of registers that equal 0.
When $n^{(i)} \ll M/w$, we easily find that $q_\text{R}^{(i)}$ is approximately computed as the fraction of registers that equal 0 at time $i$,
and we have $\mathbb{E}(\frac{1}{q_\text{B}^{(i)}}) < \mathbb{E}(\frac{1}{q_\text{R}^{(i)}})$ because  the number of bits in FreeBS is $w$ times larger than the number of registers in FreeRS under the same memory usage.
It indicates that FreeBS is more accurate than FreeRS for users whose user-item pairs appear early in stream $\Gamma$.

\section{Evaluation} \label{sec:results}
\subsection{Datasets}
\begin{figure}[tb!]
\centering
\includegraphics[width=0.42\textwidth]{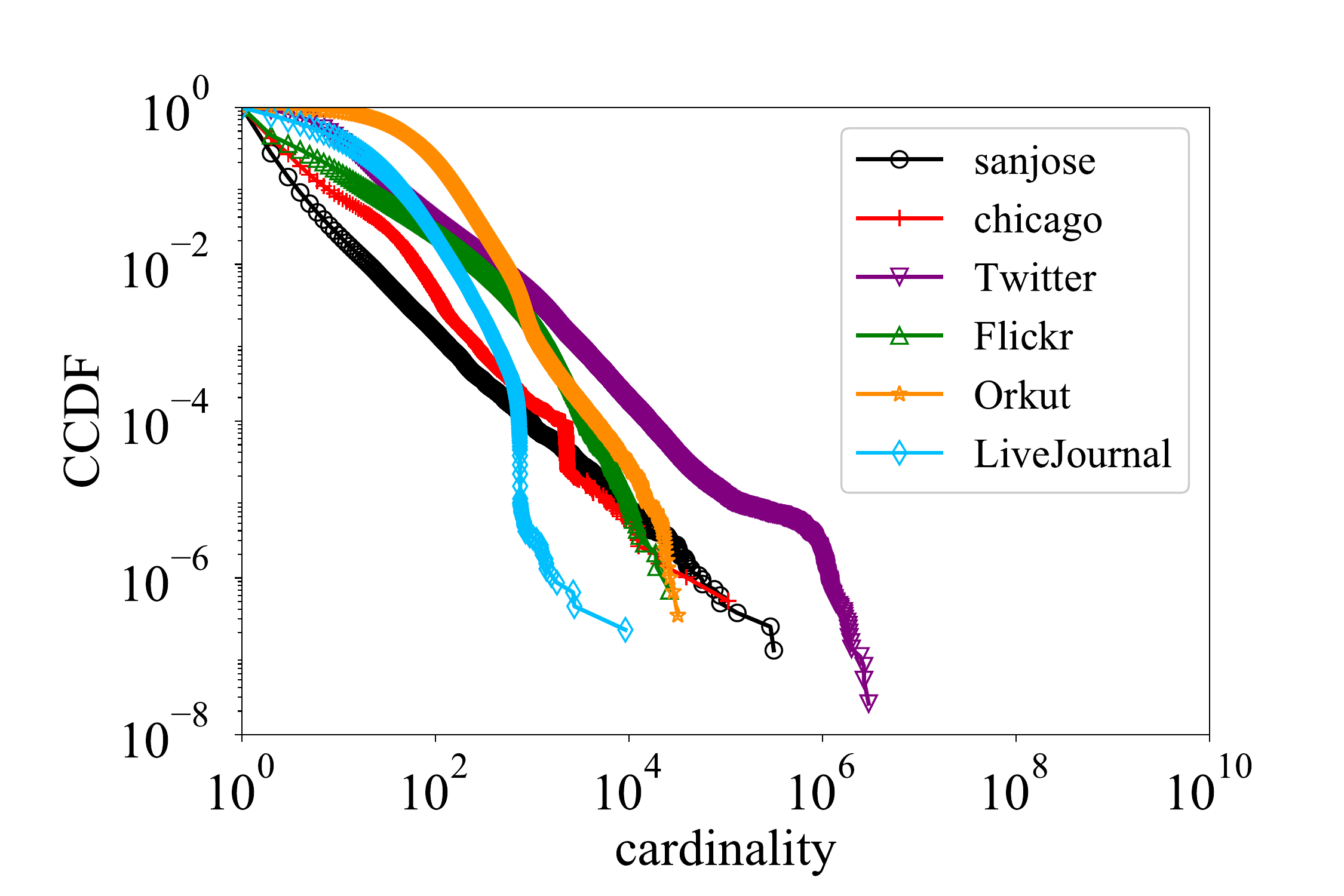}
\caption{CCDFs of user cardinalities.}
\label{fig:CCDF}
\end{figure}

In our experiments, we used a variety of publicly available real-world datasets to evaluate the performance of our methods in comparison with state-of-the-art methods,
which are summarized in Table~\ref{tab:datasets}.
Dataset sanjose (resp. chicago) consists of one-hour passive traffic traces collected from the equinix-sanjose (resp. equinix-chicago) data collection monitors during March 20, 2014.
Twitter, Flickr, Orkut and LiveJournal are all graph-datasets where each edge represents social relationship between any two users and may occur more than once in these datasets.
Figure~\ref{fig:CCDF} shows the CCDFs of user cardinalities for all datasets used in our experiments.

\begin{table}[htb]
\centering
\caption{Summary of datasets used in our experiments.
\label{tab:datasets}}
\begin{tabular}{|c|c|c|c|c|}
\hline
{\bf dataset}&{\bf $\#$users}&{\bf max-cardinality}&{\bf total cardinality}\\
\hline
sanjose~\cite{CAIDAdata} & 8,387,347 & 313,772 &23,073,907 \\
\hline
chicago~\cite{CAIDAdata} & 1,966,677 & 106,026 & 9,910,287 \\
\hline
Twitter~\cite{Kwak2010} & 40,103,281 & 2,997,496 & 1,468,365,182 \\
\hline
Flickr~\cite{MisloveIMC2007} & 1,441,431 & 26,185 & 22,613,980 \\
\hline
Orkut~\cite{MisloveIMC2007} & 2,997,376 & 31,949 & 223,534,301 \\
\hline
LiveJournal~\cite{MisloveIMC2007} & 4,590,650 & 9,186 & 76,937,805 \\
\hline
\end{tabular}
\end{table}

\subsection{Baselines}
FreeBS and CSE~\cite{Yoon2009} are bit-sharing algorithms,
while FreeRS and vHLL~\cite{XiaoSIGMETRICS2015} are register-sharing algorithms,
where each register consists of $5$ bits, i.e., $w=5$.
To compare our methods with these two methods under the same memory size,
we let FreeBS and CSE have $M$ bits,
whereas FreeRS and vHLL have $M/5$ 5-bit registers.
Moreover, both CSE and vHLL use virtual sketches to record the cardinality of each user,
and we set $m$ (i.e., the number of bits/registers in the virtual sketch) to be the same for both methods.
LPC~\cite{Whang1990} and HyperLogLog++~\cite{HeuleEDBTICDT2013} (short for HLL++) build a sketch for each user $s \in S$ to record items that $s$ connects to.
Specially, HLL++ is an optimized method of HyperLogLog,
which uses $6$ bits for each register,
and implements bias correction and sparse representation strategies to improve cardinality estimation performance.
In our experiments, under the same memory size $M$,
we let LPC have $\frac{M}{|S|}$ bits and HLL++ have $\frac{M}{6|S|}$ 6-bit registers for each user respectively.

In this paper, we aim to compute the cardinalities of all users at each time $t=1,2,\ldots$.
At each time $t$,
enumerating each occurred user and computing its cardinality requires time $O(|S^{(t)}|m)$ for methods CSE, vHLL, LPC, and HLL++,
where $S^{(t)}$ is the set of occurred users before and include time $t$.
It is computationally intensive and thus is prohibitive for estimating all users' cardinalities over time $t$.
To solve this problem, we allocate each occurred user $s$ a counter $\hat n_s$ to keep tracking of $u$'s cardinality for CSE, vHLL, LPC, and HLL++.
For each edge $e^{(t)}=(s^{(t)}, d^{(t)})$ arriving at time $t$, we only estimate the cardinality of user $s^{(t)}$ for CSE, vHLL, LPC, and HLL++, and then update its counter $\hat n_{s^{(t)}}$ and keep the counters of the other occurred users unchanged,
which reduces time complexity from $O(|S^{(t)}|m)$ to $O(m)$.
To track all users' cardinalities over time,
therefore, all methods FreeRS, FreeBS, CSE, vHLL, LPC, and HLL++ require a counter for each user,
and we do not consider this memory usage in our comparisons.

\subsection{Metrics}
In our experiments, we use a fine-grained metric \emph{relative standard error} (RSE) to evaluate the performance of estimating the cardinality for any user with a particular cardinality $n$ at time $t$.
Smaller RSE indicates better performance of cardinality estimation.
Formally, we define
\[
\text{RSE}^{(t)} (n) = \frac{1}{n} \sqrt{\frac{\sum_{s \in S}(\hat n_s^{(t)} - n)^2\mathbf{1}(n_s^{(t)}=n)}{\sum_{s \in S}\mathbf{1}(n_s^{(t)}=n)}}.
\]

\subsection{Runtime}
\begin{figure}[tb!]
\centering
\includegraphics[width=0.42\textwidth]{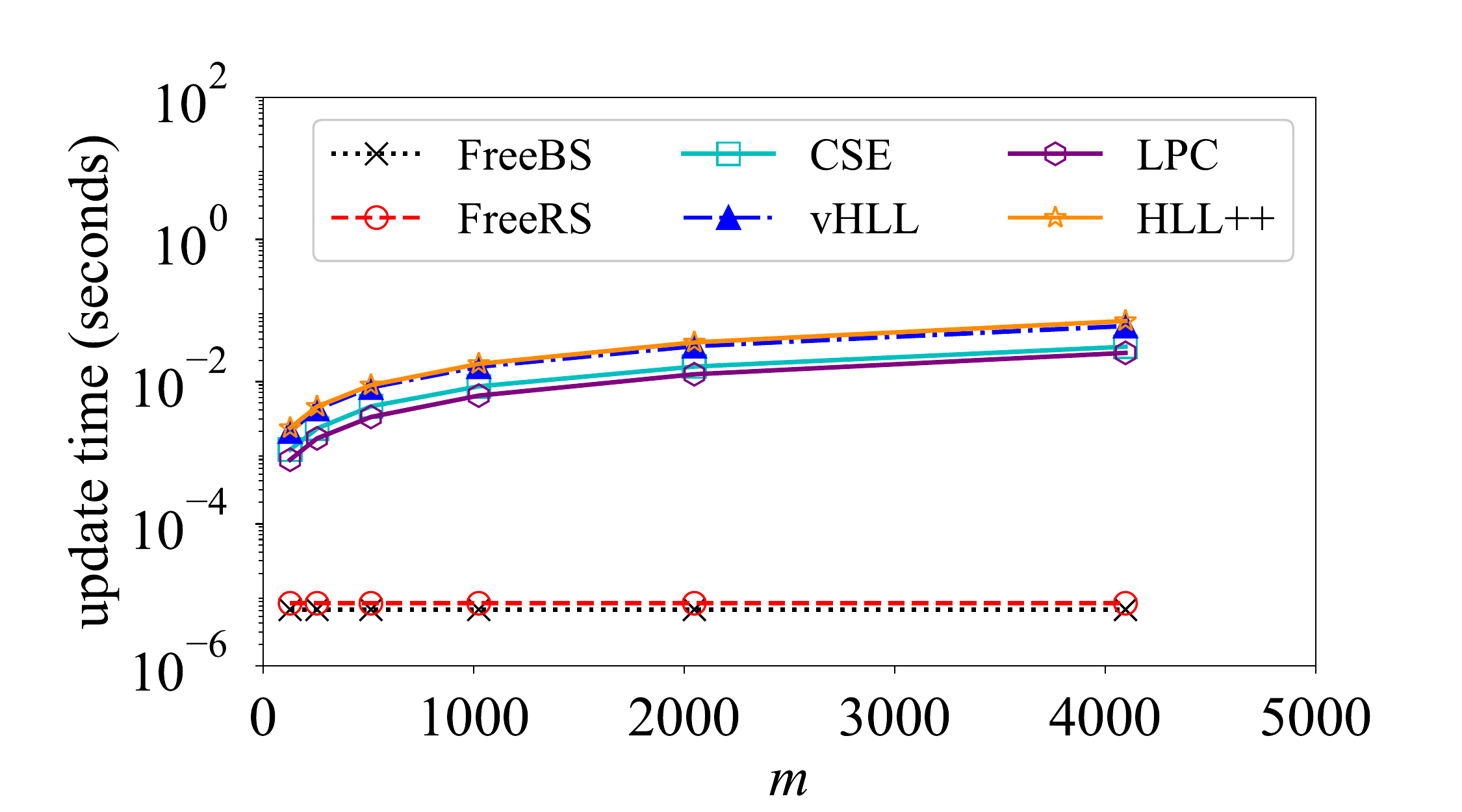}
\caption{Runtime of our methods FreeBS and FreeRS in comparison with CSE, vHLL, LPC and HLL++ for different $m$ (i.e., the number of bits/registers in the (virtual) sketch) of a user.}
\label{fig:runningtime}
\end{figure}

For CSE, vHLL, LPC and HLL++, its runtime is mainly determined by the number of bits/registers in each user's (virtual) sketch,
because estimating a user's cardinality needs to enumerate $m$ bits/registes in the user's sketch,
which requires large time complexity $O(m)$.
On the contrary, our methods FreeBS and FreeRS have low time complexity $O(1)$ to process each coming element and update the counter for its user.
In our experiments, we vary the number of bits/registers in the sketch $m$ to compare the runtime of all six methods.
In this case, we record the runtime required for processing each element and updating the cardinality of the user,
and then average the update time for all users in the datasets.
The experimental results are shown in Figure~\ref{fig:runningtime}.
As $m$ increases, the runtime of all four methods CSE, vHLL, LPC and HLL++ increases.
Our methods FreeBS and FreeRS are substantially faster than other four methods for different $m$.
Also we notice that CSE is faster than vHLL and FreeBS is faster than FreeRS, this is because register sharing based methods perform more operations than bitmap sharing methods for processing each element.

\subsection{Accuracy}
\begin{figure}[tb!]
\centering
\subfigure[FreeBS]{\includegraphics[width=0.23\textwidth]{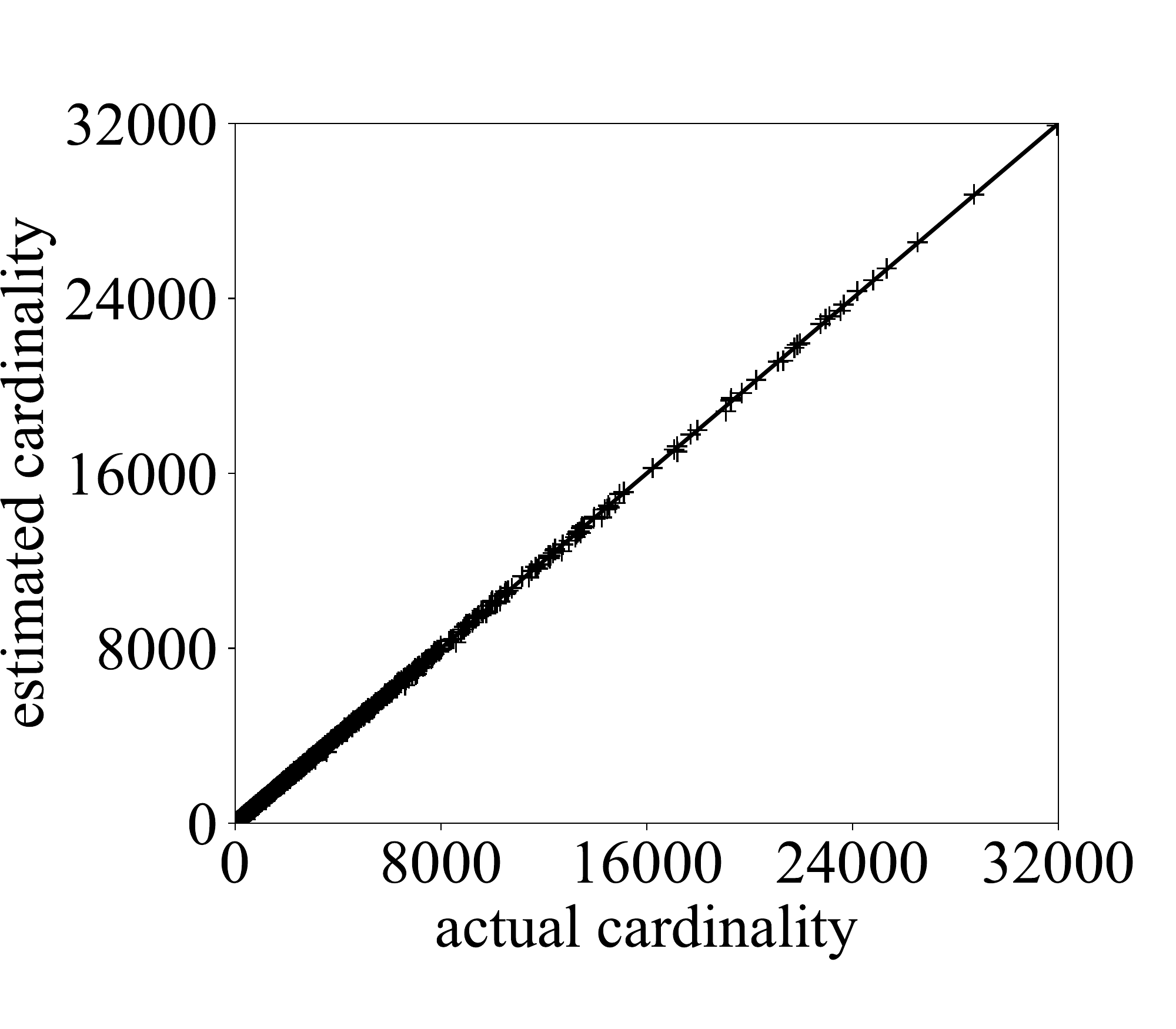}}
\subfigure[FreeRS]{\includegraphics[width=0.23\textwidth]{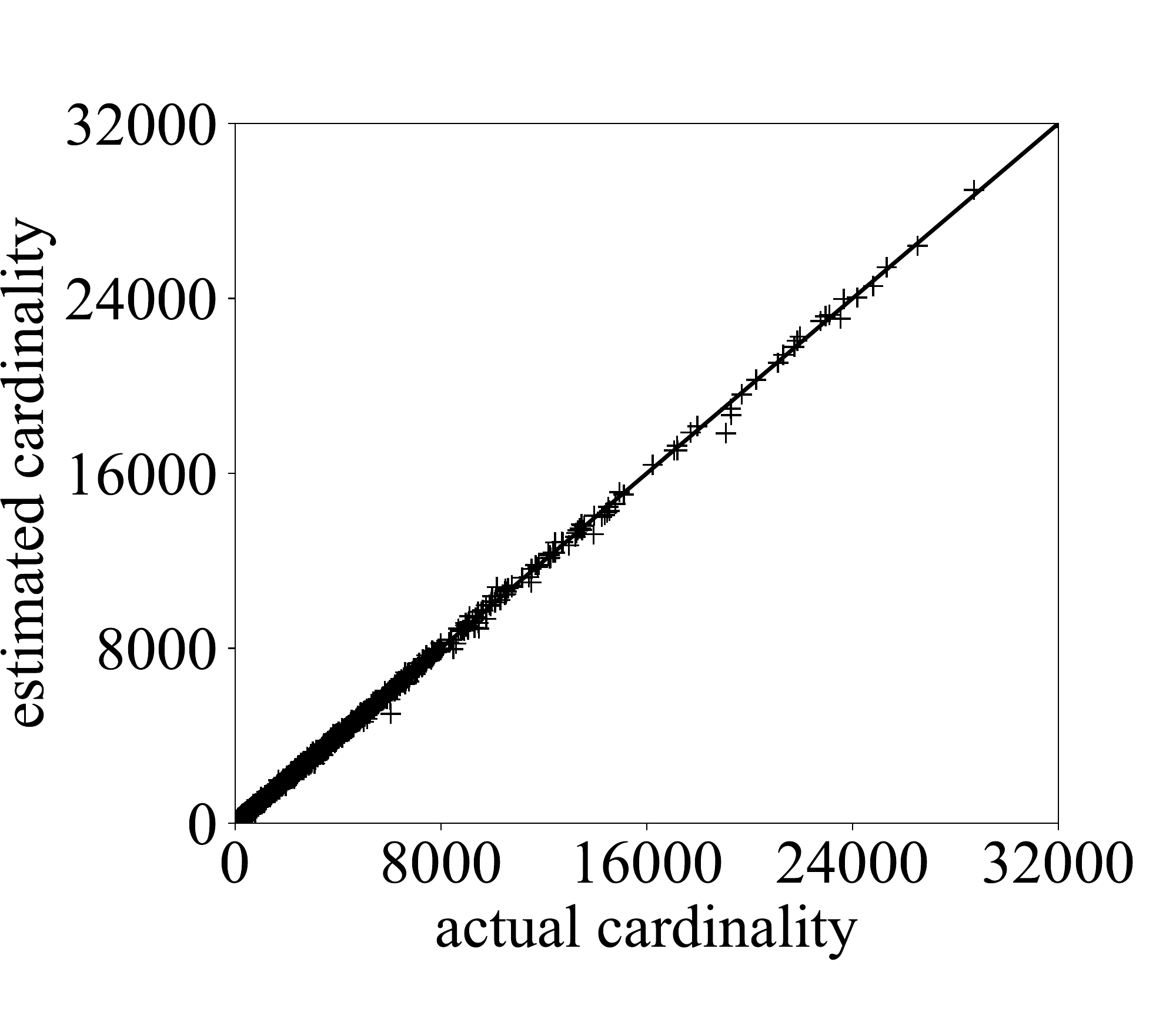}}
\subfigure[CSE]{\includegraphics[width=0.23\textwidth]{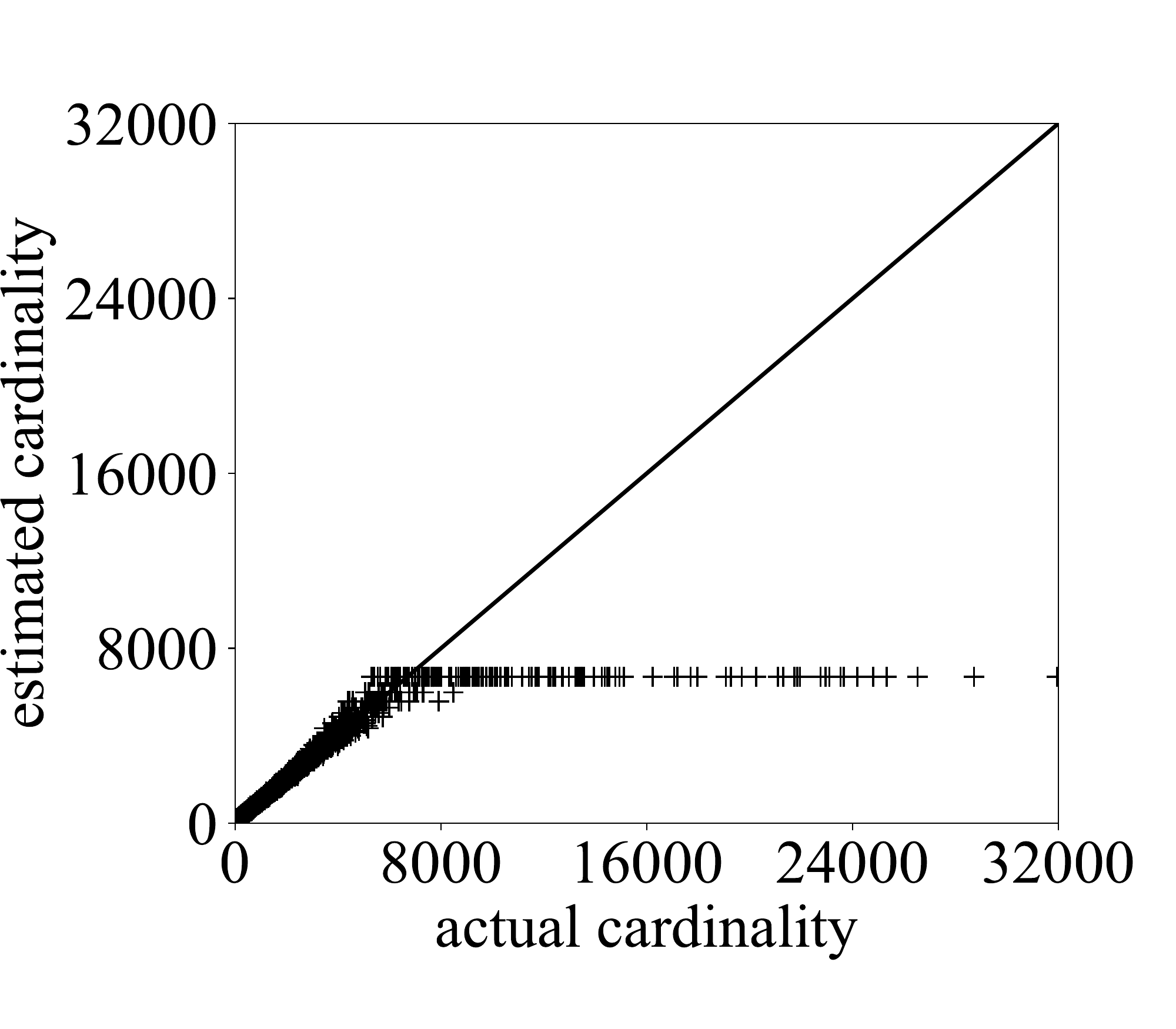}}
\subfigure[vHLL]{\includegraphics[width=0.23\textwidth]{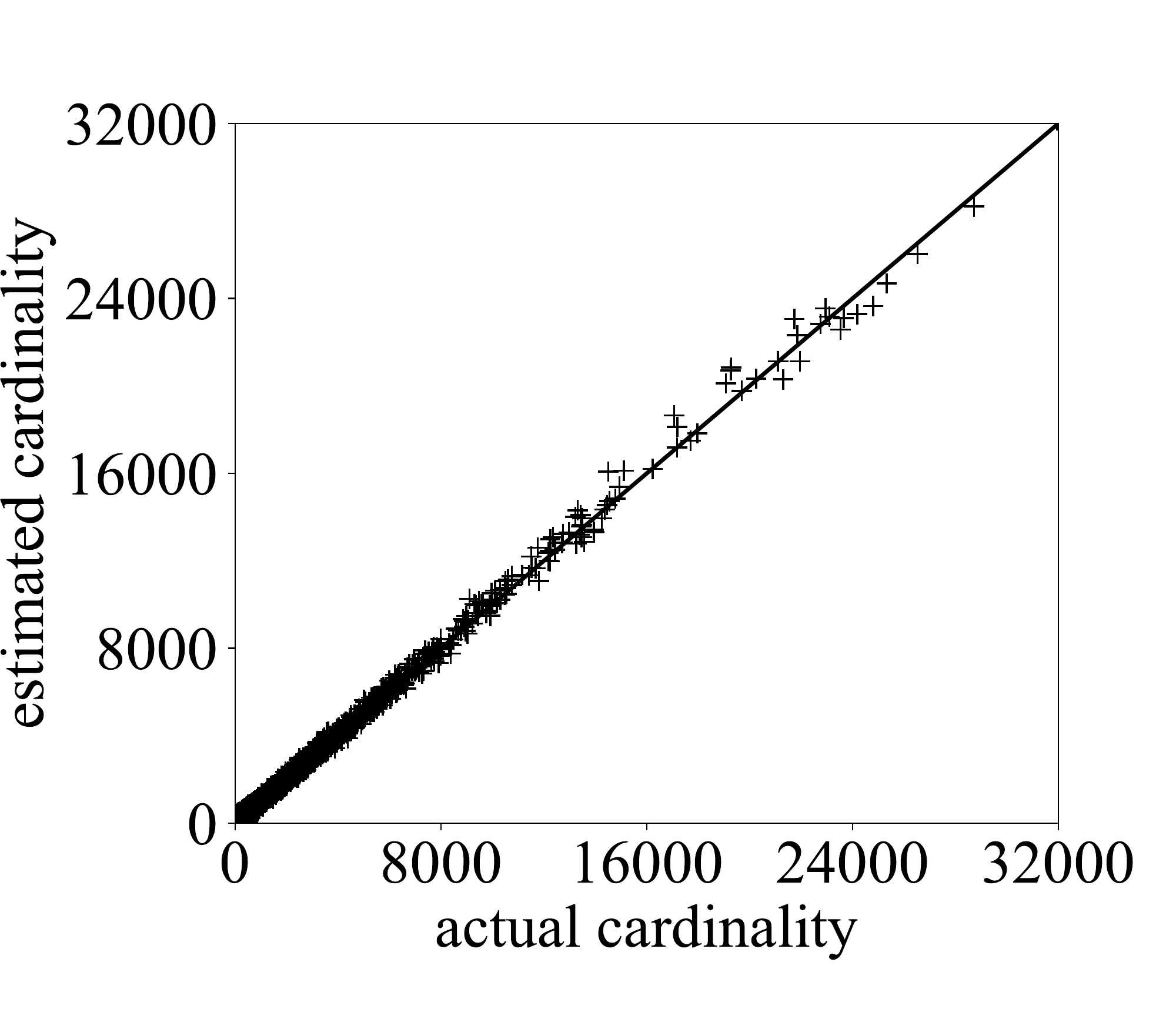}}
\subfigure[LPC]{\includegraphics[width=0.23\textwidth]{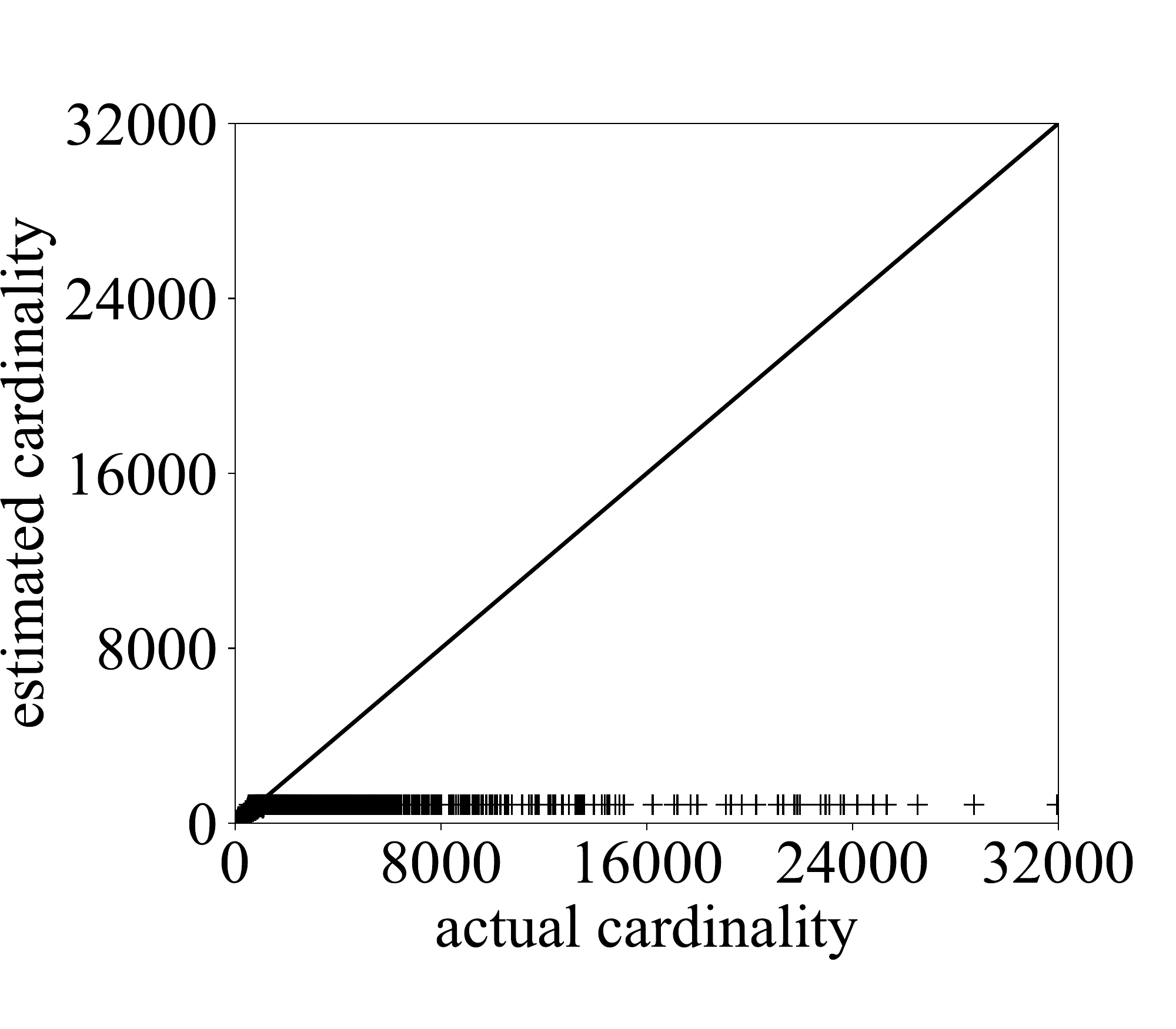}}
\subfigure[HLL++]{\includegraphics[width=0.23\textwidth]{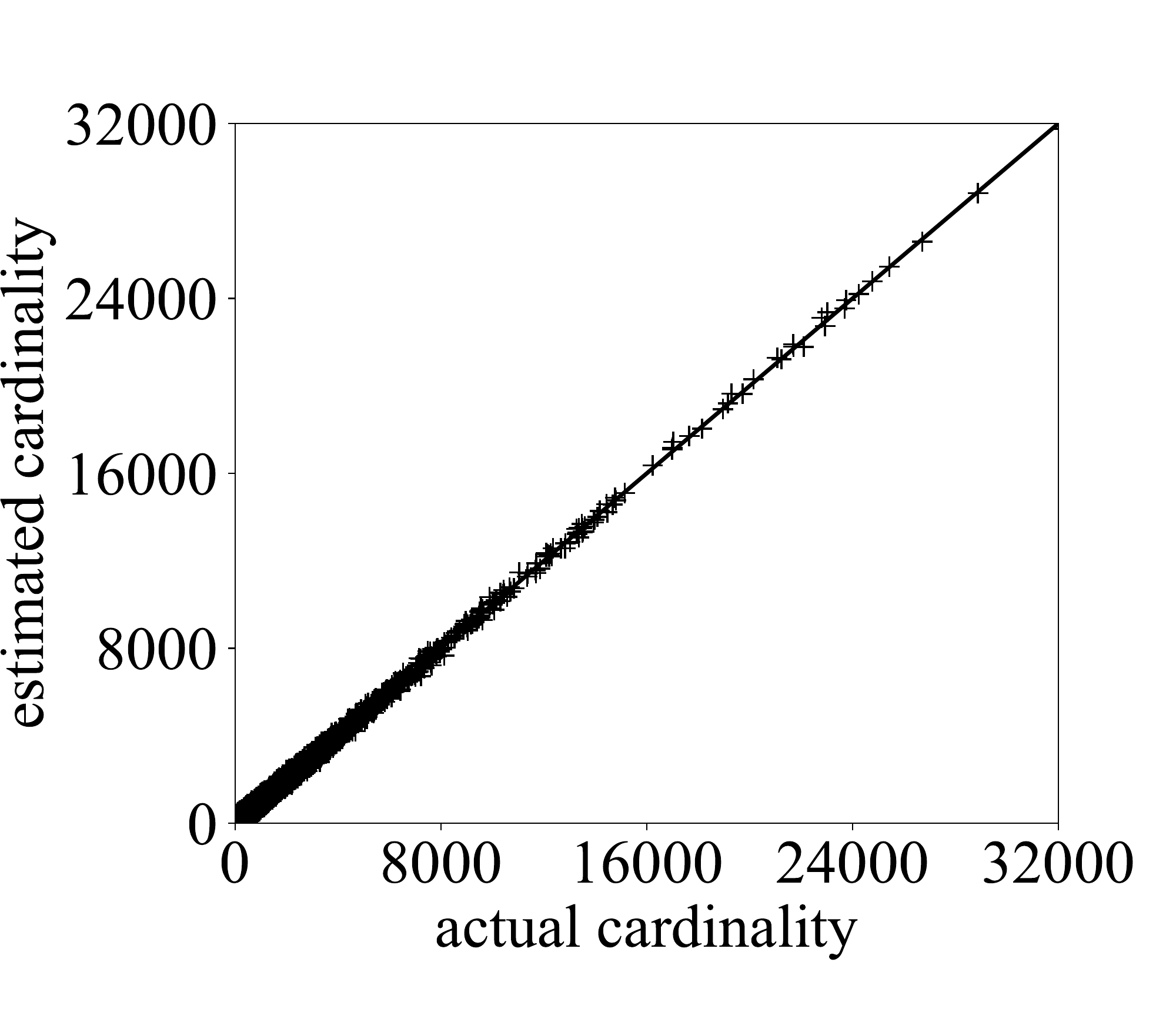}}
\caption{(Orkut) Estimated cardinalities vs actual cardinalities for FreeBS, FreeRS, CSE, vHLL, LPC, and HLL++ under the same memory size $M=5 \times 10^{8}$ (bits), where the number of bits/registers in the virtual sketch for CSE/vHLL $m=1,024$.}
\label{fig:scatter}
\end{figure}

\begin{figure*}[htb]
\centering
\subfigure[sanjose]{\includegraphics[width=0.3\textwidth]{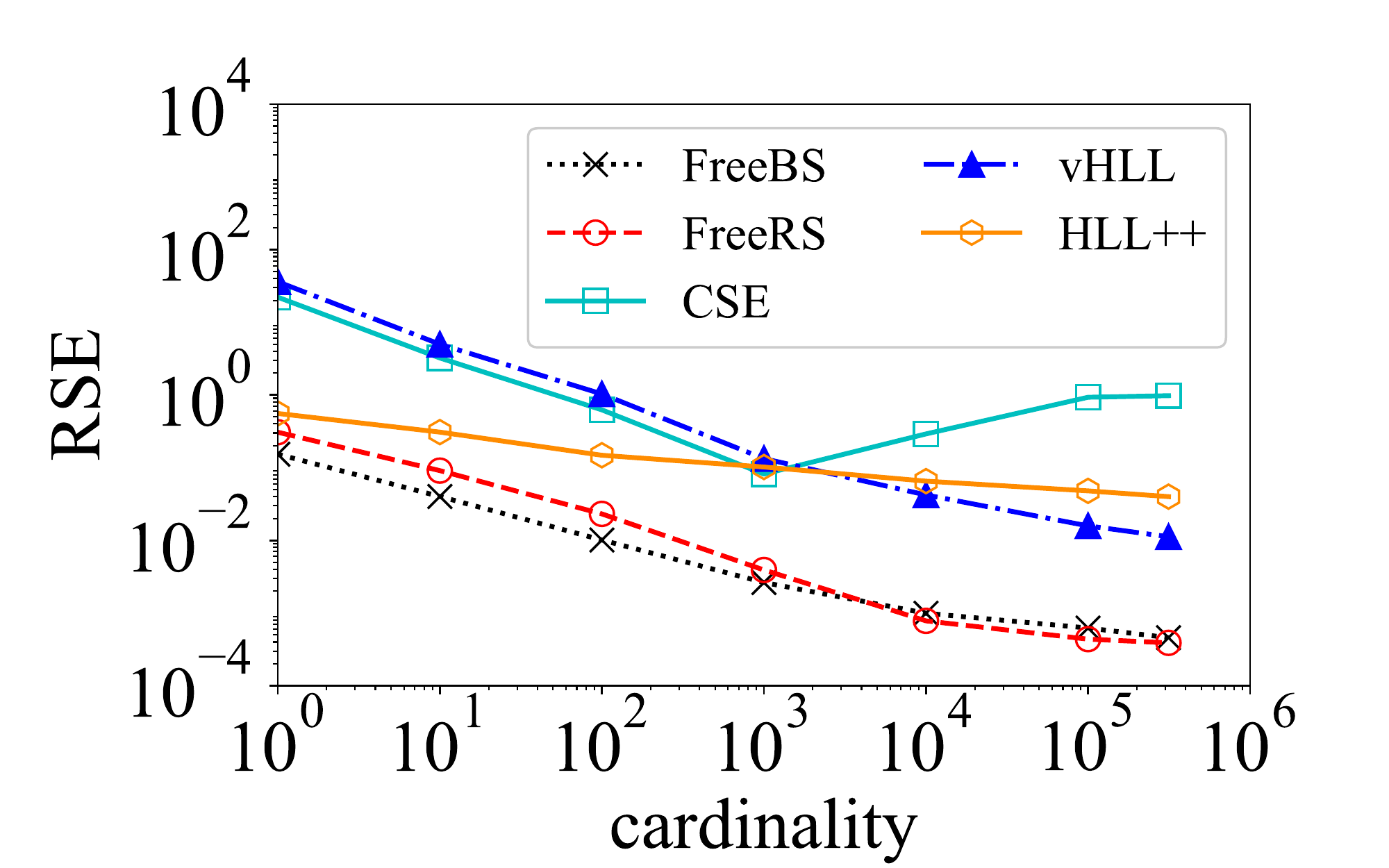}}
\subfigure[chicago]{\includegraphics[width=0.3\textwidth]{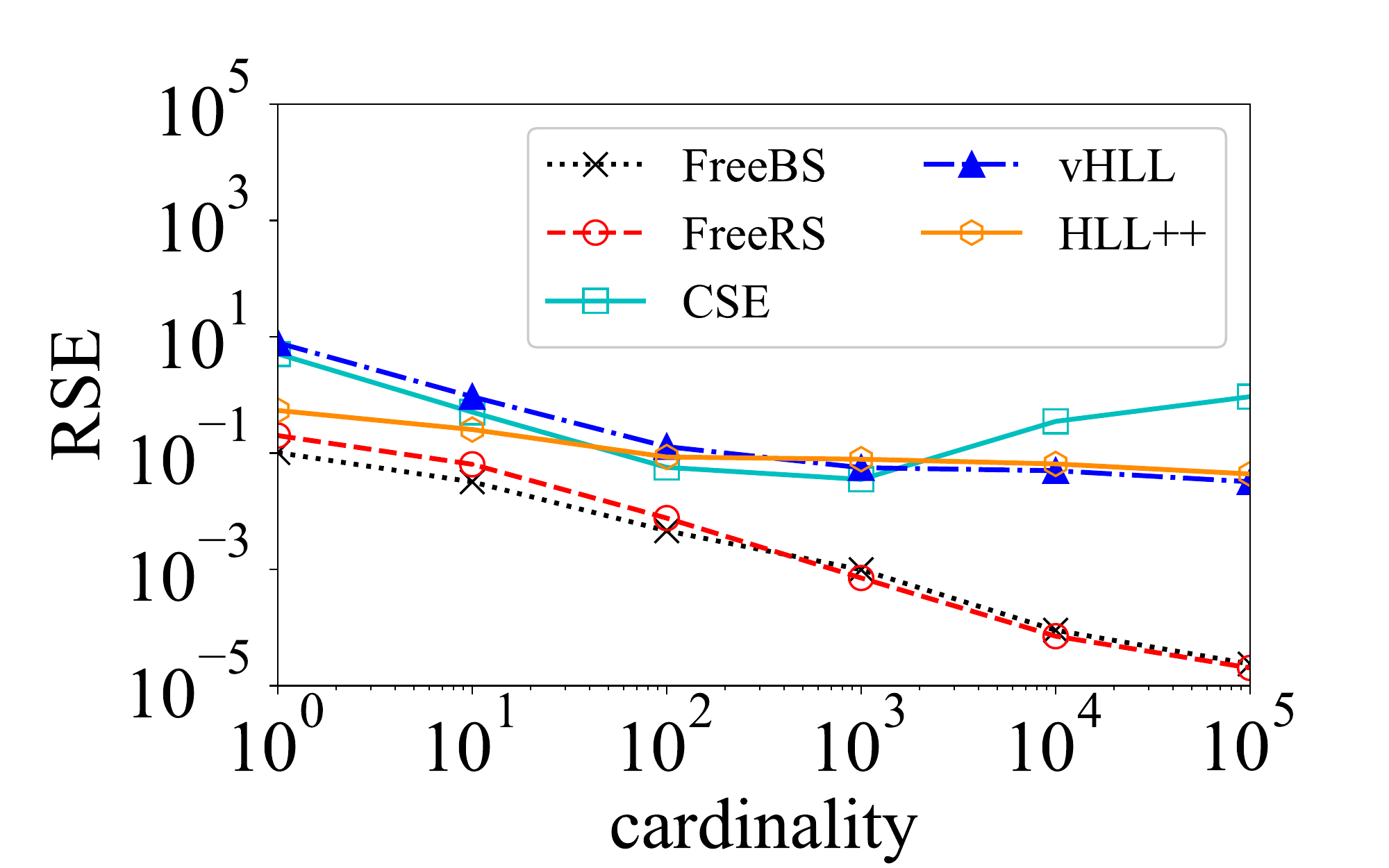}}
\subfigure[Twitter]{\includegraphics[width=0.3\textwidth]{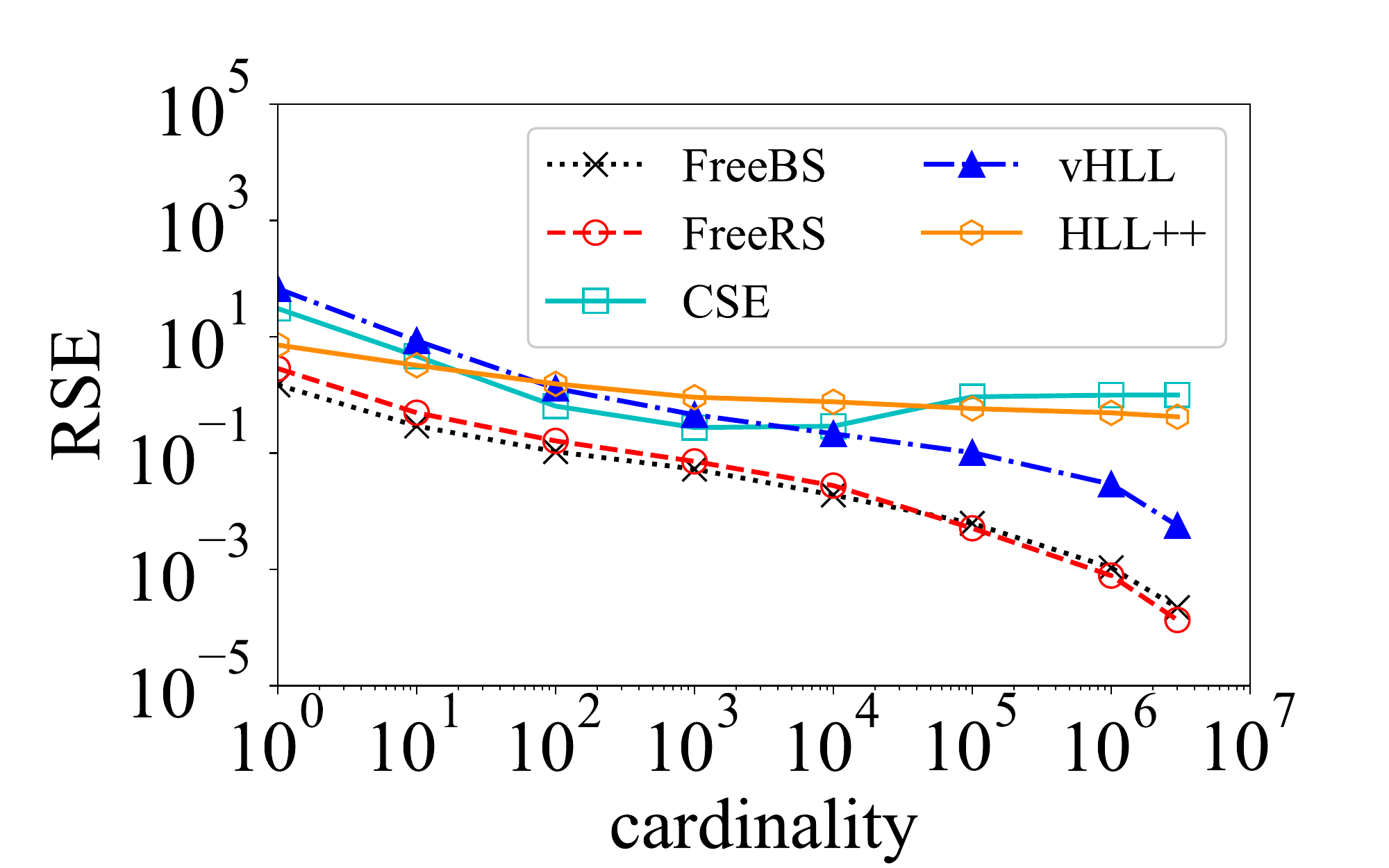}}
\subfigure[Flickr]{\includegraphics[width=0.3\textwidth]{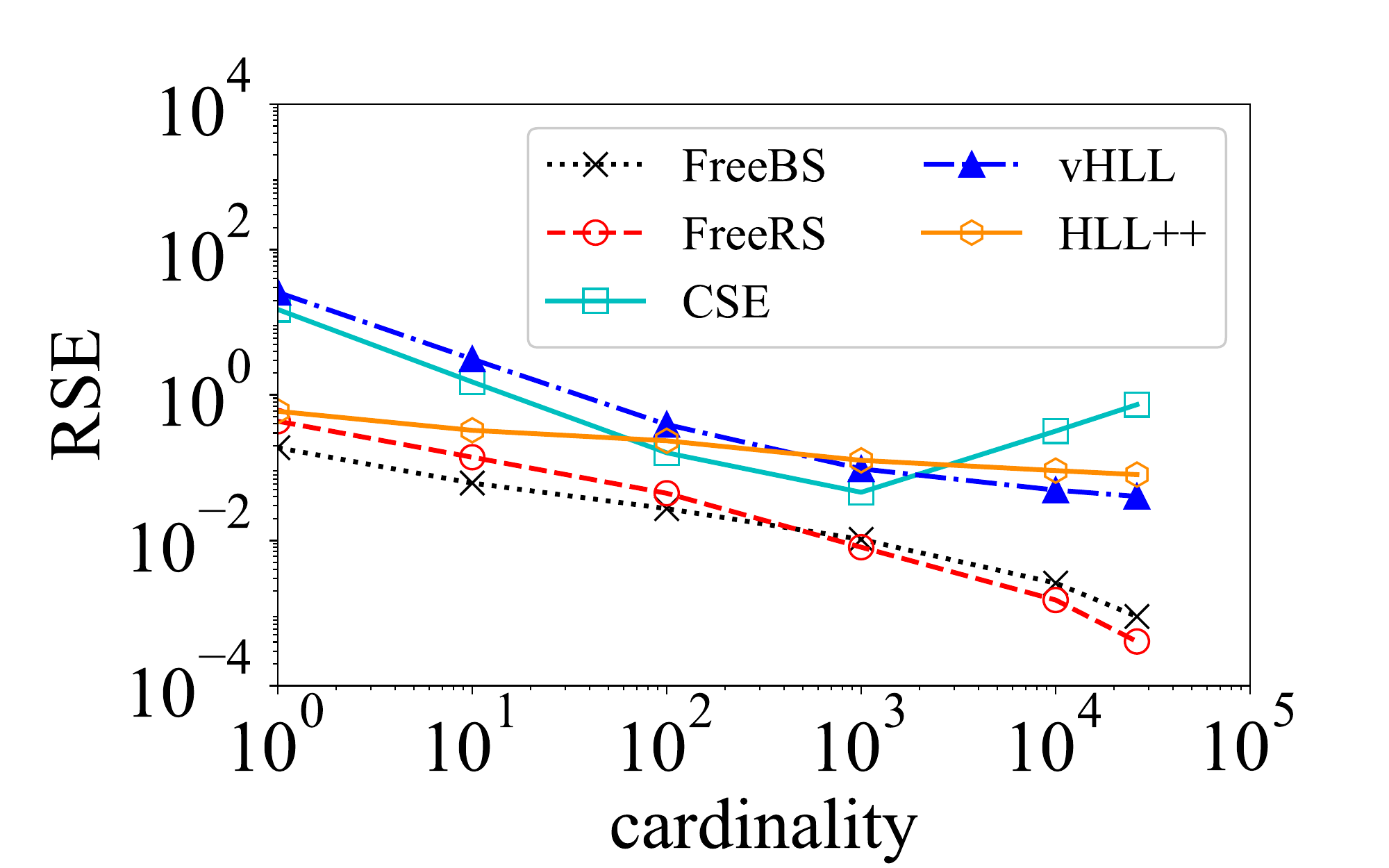}}
\subfigure[Orkut]{\includegraphics[width=0.3\textwidth]{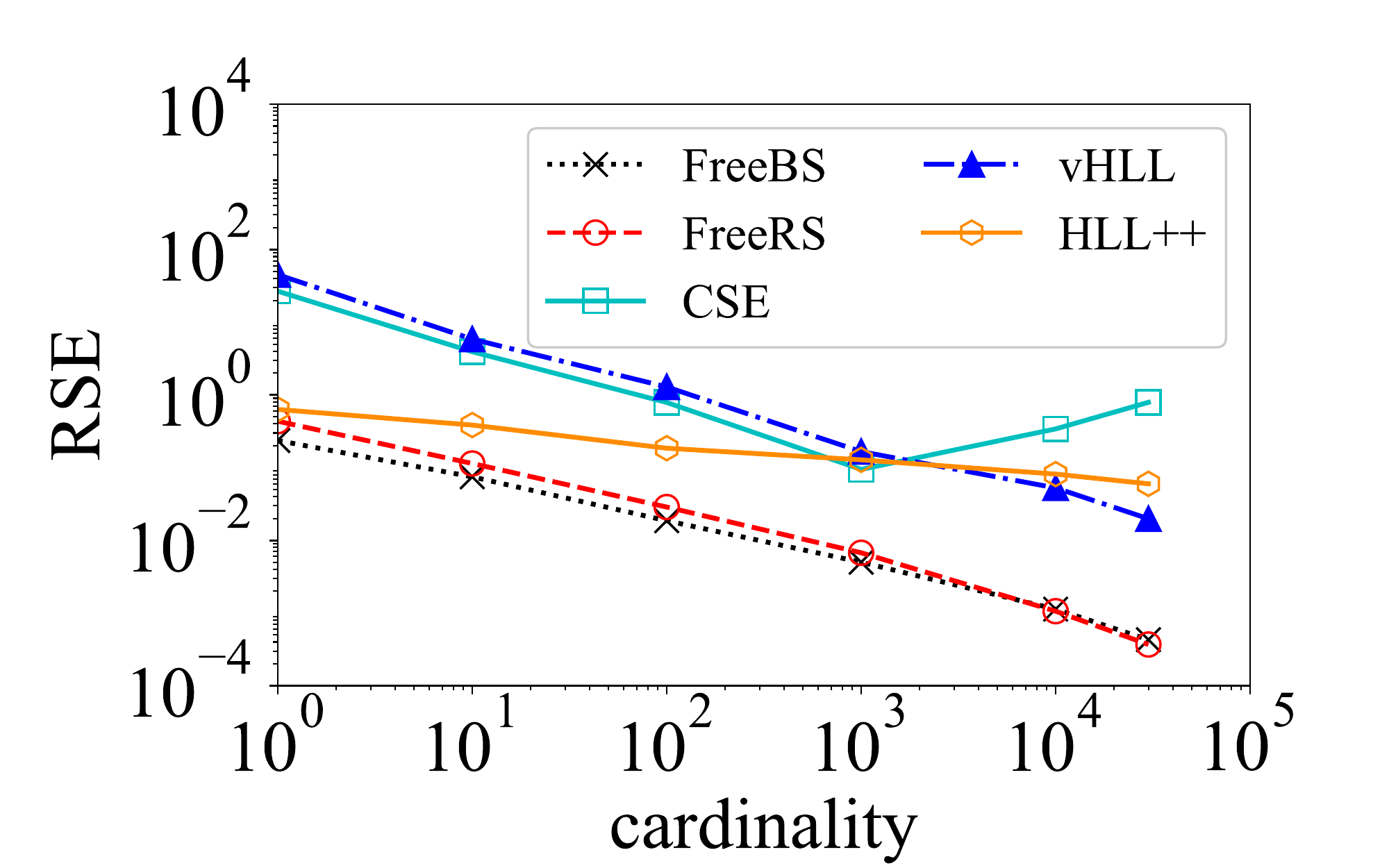}}
\subfigure[LiveJournal]{\includegraphics[width=0.3\textwidth]{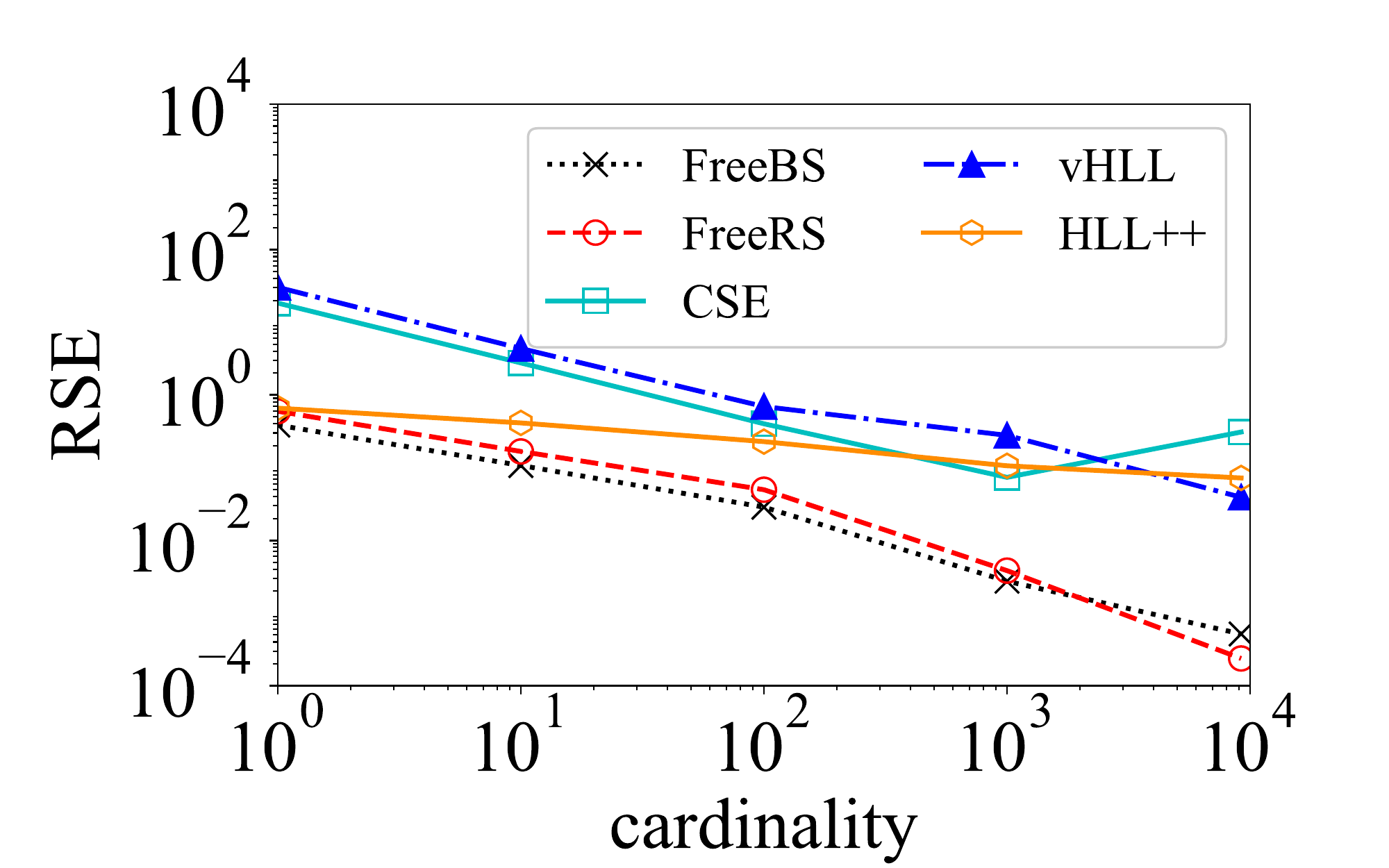}}
\caption{(All datasets) Cardinality estimation accuracy, where memory size $M=5 \times 10^{8}$ (bits), and $m=1,024$ for CSE/vHLL.}
\label{fig:RSE}
\end{figure*}

Whenever an element $e=(s,d)$ arrives, we implemented all six methods to estimate the cardinality of the user $s$.
Figure~\ref{fig:scatter} shows the experimental results when all elements in the dataset Orkut arrive.
In our experiments, we fixed the memory size $M=5 \times 10^8$ bits,
and the number of bits/register in the virtual sketch for CSE/vHLL was set to $m=1,024$.
Under the fixed memory size, each user in the dataset Orkut used $167$ bits for LPC and $28$ 6-bit registers for HLL++ respectively.
Points close to the diagonal line $\hat n_s^{(t)} = n_s^{(t)}$ indicate better cardinality estimation accuracy.
From Figure~\ref{fig:scatter}, we observe that our methods FreeBS and FreeRS are more accurate than other four methods for different actual cardinalities.
CSE and LPC have limited estimation ranges and LPC exhibits extremely bad estimation performance especially for larger cardinalities,
therefore we omit the experimental results of LPC in the later experiments.
Furthermore, we used the metric RSE to compare the estimation errors of our methods with those of CSE, vHLL and HLL++ at a fine-grained level.
Figure~\ref{fig:RSE} shows the RSEs of all methods for all datasets in Table~\ref{tab:datasets}.
We can see that
our methods FreeBS and FreeRS are up to 10,000 times more accurate than CSE, vHLL, and HLL++.
Bit sharing method FreeBS (resp. CSE) is more accurate than register sharing method FreeRS (resp. vHLL) for characterizing users with small cardinalities.
For users with large cardinalities, register sharing method FreeRS (resp. vHLL) outperforms bit sharing method FreeBS (resp. CSE).
Specially, the RSE of CSE first decreases and then increases as the actual cardinality increases.
This is because CSE has a small estimation range, i.e., $m \ln m$,
which is consistent with the original paper~\cite{Yoon2009}.
Meanwhile, HLL++ is more accurate than CSE and vHLL for small cardinalities due to its bias correction strategy.
Because each user uses fewer registers for HLL++ than vHLL, HLL++ performs larger estimation errors than vHLL for large cardinalities.

\begin{figure*}[tb!]
\centering
\subfigure{\includegraphics[width=0.42\textwidth]{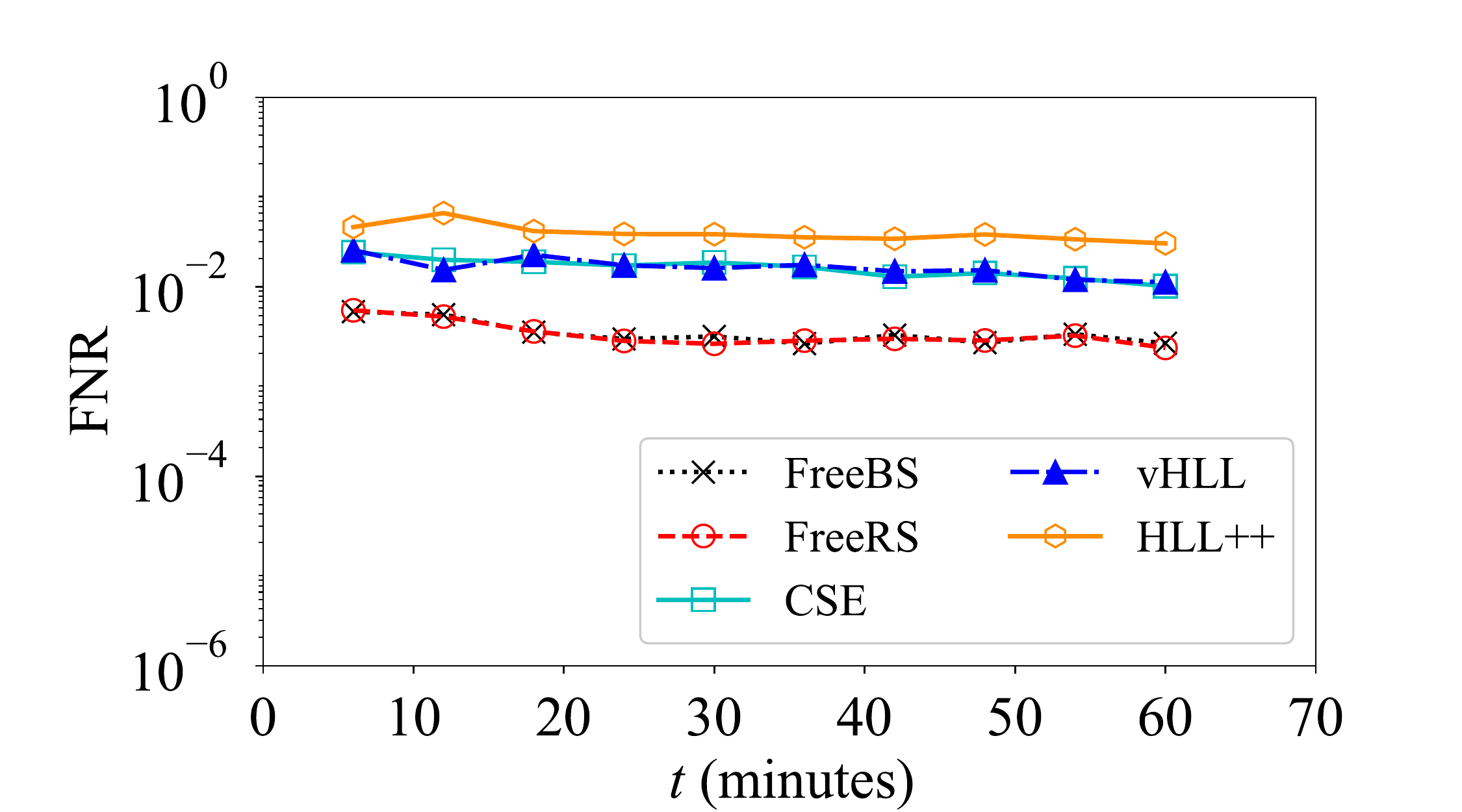}}
\subfigure{\includegraphics[width=0.42\textwidth]{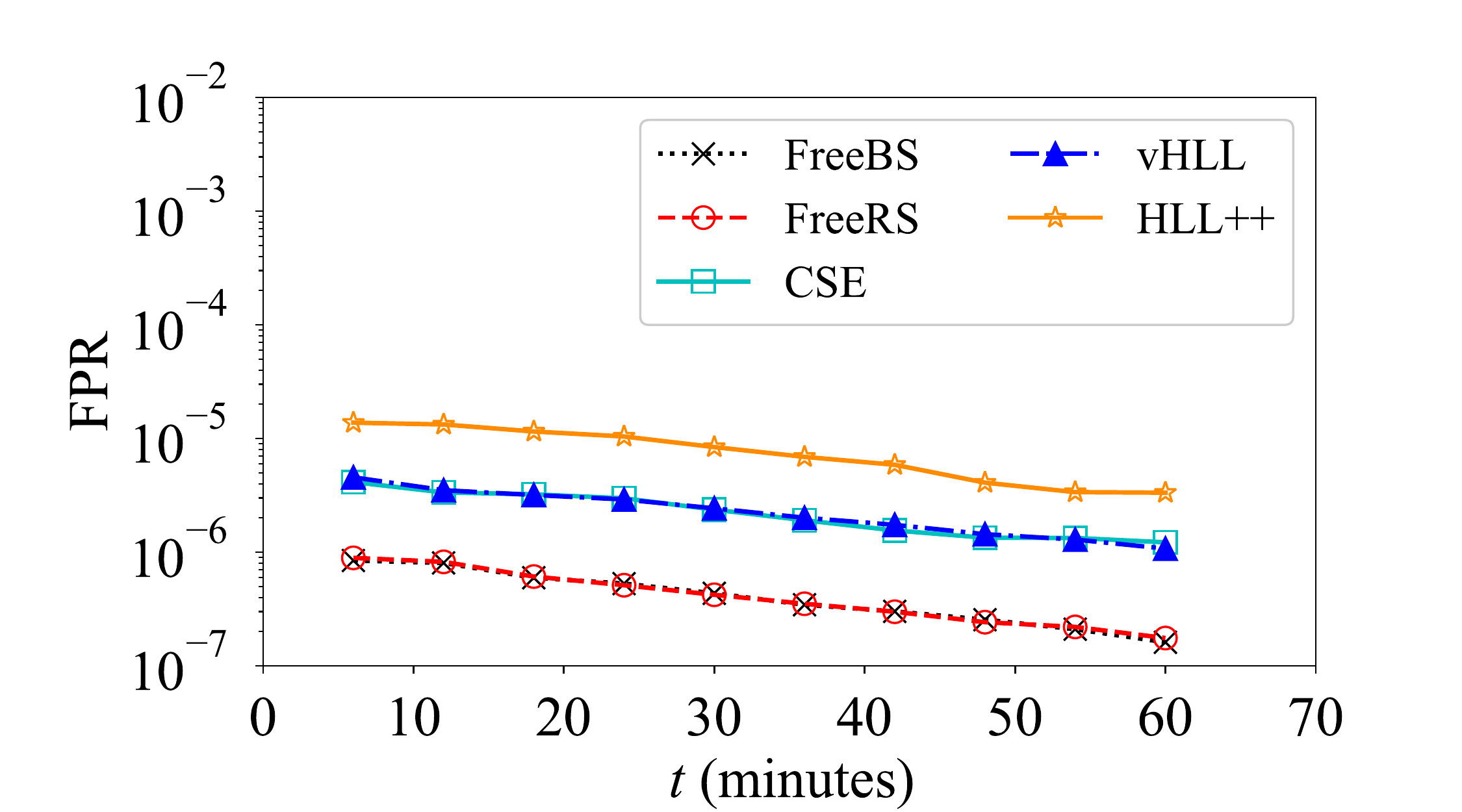}}
\caption{(sanjose) Accuracy of detecting super spreaders over time $t$,
where $\Delta=5 \times 10^{-5}$, $M=5 \times 10^8$ (bits), and $m=1,024$ for CSE/vHLL.}
\label{fig:Superspreader}
\end{figure*}

\begin{table*}[tb!]
	\caption{(All datasets) Performance of detecting super spreaders with $\Delta=5 \times 10^{-5}$, $M=5 \times 10^{8}$, $m=1,024$. CSE reported an empty set of users for Twitter and Orkut due to the limited estimation range, therefore we report their results on Twitter and Orkut as ``N/A".}
	\label{tab:Tabsuperspreader}
	\centering
	\begin{tabular}{|c|c|c|c|c|c|c|c|c|c|c|}
		\hline
		\multirow{2}{*}{\textbf{dataset}}&\multicolumn{5}{c|}{\textbf{FNR}} & \multicolumn{5}{c|}{\textbf{FPR}} \\
        \cline{2-11}
		& \textbf{FreeBS} & \textbf{FreeRS} & \textbf{CSE} & \textbf{vHLL} & \textbf{HLL++} & \textbf{FreeBS} & \textbf{FreeRS} & \textbf{CSE} & \textbf{vHLL} & \textbf{HLL++} \\
        \hline
		sanjose & 2.54e-3 & \textbf{2.27e-3} & 1.02e-2 & 1.12e-2 & 2.88e-2 & \textbf{1.61e-7} & 1.76e-7 & 1.22e-6 & 1.08e-6 & 3.38e-6 \\
        \hline
		chicago & 2.61e-3 & \textbf{2.55e-3} & 8.00e-3 & 8.23e-3 & 9.06e-3 & \textbf{2.32e-6} & 2.54e-6 & 8.46e-6 & 8.47e-6 & 1.08e-5 \\
        \hline
		Twitter & 2.45e-2 & \textbf{2.27e-2} & N/A & 6.38e-2 & 1.04e-1 & 2.99e-7 & \textbf{2.75e-7} & N/A & 6.84e-7 & 1.25e-6 \\
        \hline
		Flickr & \textbf{5.13e-3} & 5.55e-3 & 1.09e-2 & 1.13e-2 & 1.38e-2 & 1.18e-5 & \textbf{1.04e-5} & 3.57e-5 & 3.87e-5 & 4.13e-5 \\
        \hline
		Orkut & \textbf{1.47e-2} & 1.53e-2 & N/A & 7.86e-2 & 9.71e-2 & \textbf{3.22e-7} & 3.34e-7 & N/A & 1.97e-6 & 2.52e-6 \\
        \hline
		LiveJournal & \textbf{4.37e-3} & 4.62e-3 & 1.04e-2 & 9.92e-3 & 1.36e-2 & 4.33e-7 & \textbf{4.15e-7} & 1.22e-6 & 1.13e-6 & 1.78e-6 \\
        \hline
	\end{tabular}
\end{table*}

\subsection{Case Study: Detecting Super Spreaders over Time}
We implemented  methods FreeBS, FreeRS, CSE, vHLL, and HLL++ to detect super spreaders over time,
where a super spreader refers to a user connecting to at least $\Delta n^{(t)}$ items at time $t$,
where $n^{(t)}$ is the sum of all user cardinalities at time $t$ and $0<\Delta<1$ is a relative threshold.
In our experiments, we set memory size $M = 5 \times 10^8$ bits and the number of bits/registers in the virtual sketch $m = 1,024$.
We used two metrics \emph{false negative ratio} (FNR) and \emph{false positive ratio} (FPR) to evaluate performance,
where FNR is the ratio of the number of super spreaders not detected to the number of super spreaders,
and FPR is the ratio of the number of users that are wrongly detected as super spreaders to the number of all users.
Figure~\ref{fig:Superspreader} shows experimental results of all five methods for detecting super spreaders over time in dataset sanjose.
We observe that both FreeBS and FreeRS are more accurate than other three methods for detecting super spreaders over time.
For example, FNR and FPR for FreeBS and FreeRS are about $4$ to $20$ times smaller than other three methods.
Table~\ref{tab:Tabsuperspreader} shows the results for all datasets when all elements arrive.
We notice that our methods FreeBS and FreeRS outperform other three methods on all datasets.

\section{Related Work} \label{sec:related}
Sketch methods use small amount of memory to quickly build compressive summaries of large data streams,
and have been successfully used for applications such as heavy hitter detection~\cite{Estan2002,EstanSigcomm2003,CormodeVLDB2003,CormodeSigmod2004,Zhang2004,YuNSDI2013}, heavy change detection~\cite{Krishnamurthy2003,Cormode2003,Schweller2007}, super host detection~\cite{Zhao2005,Cao2009,Yoon2009,Wang2011,WangTIFS2012,WangCN2012}, cardinality distribution estimation~\cite{ChenINFOCOM2009,WangTIFS2014}, network flow size estimation~\cite{Duffield2003,Hohn2003,Yang2007,KumarSIGMETRICS2004,Ribeiro2008,Kumar2005,Kumar2006,LievenS10}, and network traffic entropy estimation~\cite{Lall2006,Zhao2007}.
Next, we discuss existing cardinality estimation methods in detail.\\
\noindent\textbf{Estimating data streams' cardinalities.}
To estimate the cardinality of a large data stream (i.e., the number of distinct elements in the data stream),
Whang et al.~\cite{Whang1990} develop the first sketch method LPC.
LPC can only estimate cardinalities less than $m\ln m$,
where $m$ is the size of the LPC sketch.
Therefore, it needs to set a large $m$ to handle data streams with large cardinalities.
\cite{Estan2003,ChenJASA2011} combine LPC and different sampling methods to enlarge the estimation range.
Flajolet and Martin~\cite{Flajolet1985} develop a sketch method FM,
which uses a register to estimate the data stream's cardinality
and provides a cardinality estimation bounded by $2^w$,
where $w$ is the number of bits in the register.
To further improve the accuracy and decrease the memory usage,
sketch methods MinCount~\cite{Bar-YossefRANDOM2002}, LogLog~\cite{Durand2003}, HyperLogLog~\cite{FlajoletAOFA07}, HyperLogLog++~\cite{HeuleEDBTICDT2013}, RoughEstimator~\cite{KanePODS2010}, and HLL-TailCut+~\cite{XiaoZC17} are developed to use a list of $m$ registers
and compress the size of each register from 32 bits to 5 or 4 bits under the same estimation range of $2^{32}$.
\cite{GiroireDAM2009,lumbrosoAOFA2010} develop several cardinality estimators based on order statistics of observed samples.
Ting~\cite{TingKDD2014} introduces the concept of an area cutting process to generally model the above sketch methods and provides a martingale based estimator to further improve the accuracy of these methods.
Chen et.al~\cite{Chen2013} extend HyperLogLog to estimate the cardinality over sliding windows.
Besides these sketch methods, sampling methods Wegman's adaptive sampling~\cite{FlajoletComputing1990} and distinct sampling~\cite{GibbonsPVLDB2001} are also developed for estimating a large stream's cardinality,
while Flajolet et al.~\cite{FlajoletAOFA07} reveal that the memory efficiency of these two sampling methods is even worse than the original FM method.
Recently,~\cite{CohenKDD2017,TingKDD2016} developed methods using the above sketch methods to estimate the cardinalities of set unions and intersections.

\noindent\textbf{Estimating all user cardinalities.}
Significant attention has been paid to develop sketch methods to estimate the cardinalities of network hosts (or users) over high speed links.
To achieve a desired accuracy,
the above methods require a large number of bits/registers for each host
because host cardinalities are unknown in advance and vary over a large range.
This is not memory efficient because most hosts have small cardinalities.
To solve this problem, \cite{Zhao2005, Yoon2009, WangTIFS2012, XiaoSIGMETRICS2015} develop different virtual sketch methods to compress the LPC/HLL sketches of all hosts into a large bit/register array shared by all hosts.
Zhao et al.~\cite{Zhao2005} propose a virtual sketch method, which consists of a list of LPC sketches (i.e., a two-dimensional bit array).
For each host, they randomly select $k$ ($k$ is usually set to 2 or 3) LPC sketches from the list shared by all hosts.
Note that each LPC  in the list may be used by more than one hosts,
which introduces ``noisy" bits in a host's LPCs.
To remove the error introduced by ``noisy" bits,
Zhao et al.~\cite{Zhao2005} develop a method to estimate a host's cardinality based on its $k$ LPC sketches.
To further reduce the memory usage, \cite{Yoon2009, WangTIFS2012} generate each host's virtual LPC sketch by selecting $m$ bits from a large one-dimensional bit array at random.
All these virtual sketch methods~\cite{Yoon2009, WangTIFS2012} have to set a large value of $m$ (e.g., thousand) to achieve reasonable accuracy for host cardinalities over a large range.
It results in large estimation errors,
because most hosts have small cardinalities and many bits in their virtual LPC sketches tend to be contaminated by ``noisy" bits.
To reduce ``noisy" bits in virtual LPC sketches, \cite{TaoWGH17} builds a small regular LPC sketch for each host,
which also provides information for estimating the host cardinality.
As we mentioned, these LPC based virtual sketch methods have small estimation ranges bounded by $m\ln m$.
To enlarge the estimation range, \cite{XiaoSIGMETRICS2015} develop a sketch method vHLL,
which generates a virtual HLL sketch by randomly selecting $m$ registers from a large register array shared by all hosts.
To achieve desired accuracy for host cardinalities over a large range,
\cite{XiaoSIGMETRICS2015} also needs to set a large value of $m$.
However, it results in a high computational cost and large estimation errors for network hosts with small cardinalities,
in which virtual HLL sketches include many ``noisy" registers.
In addition, the above virtual sketch methods are customized to non-streaming settings, i.e., estimate host cardinalities at the end of an interval,
and are computationally expensive to be extended to streaming settings.

\section{Conclusions and Future Work} \label{sec:conclusions}
In this paper, we develop two novel streaming algorithms FreeBS and FreeRS to accurately estimate user cardinalities over time.
Compared to existing bit/register sharing methods using (i.e., selecting) only $m$ bits/registers for each user,
FreeBS/FreeRS enables that the number of bits/registers used by a user dynamically increases as its cardinality increases over time
and each user can use all shared bits/registers. It is therefore capable to estimate user cardinalities over a large range.
For example, existing bit sharing methods can be only used to estimate user cardinalities over the range $[0, m\ln m]$.
Our method FreeBS enlarges the range to $[0, M\ln M]$, where $M$ is the total number of bits/registers used by all users.
In addition, our algorithms FreeBS and FreeRS exploit dynamic properties of shared bits/registers to significantly improve estimation accuracy.
They are simple yet effective,
and sharply reduce time complexity of computing all user cardinalities to $O(1)$ each time they observe a new user-item pair.
We conduct experiments on real-world datasets,
and experimental results demonstrate that our methods FreeBS and FreeRS significantly outperform state-of-the-art methods
in terms of accuracy and computational time.
In future, we plan to extend our methods to applications such as SDN routers to monitor anomalies.

\section*{Acknowledgment}
The research presented in this paper is supported in part by National Key R\&D Program of China (2018YFC0830500), National Natural Science Foundation of China (U1301254, 61603290, 61602371), the Ministry of Education\&China Mobile Research Fund (MCM20160311), the Natural Science Foundation of Jiangsu Province (SBK2014021758), 111 International Collaboration Program of China, the Prospective Joint Research of Industry-Academia-Research Joint Innovation Funding of Jiangsu Province (BY2014074), Shenzhen Basic Research Grant (JCYJ20160229195940462, JCYJ20170816100819428), China Postdoctoral Science Foundation (2015M582663), Natural Science Basic Research Plan in Shaanxi Province of China (2016JQ6034).

\balance

\end{document}